\documentclass[oldversion]{aa}

\usepackage{graphicx}
\usepackage{epstopdf}

\begin{document}

\newcommand{\hi}{\ion{H}{i}~}
\newcommand{\hii}{\ion{H}{ii}~}

   \title{Evolution of spiral galaxies in modified gravity:  II-  Gas dynamics}

 \author{O. Tiret         
\and          F. Combes        
 }

   \institute{  Observatoire de Paris, LERMA, 
            61 Av. de l'Observatoire, F-75014, Paris, France   }

   \date{Received XXX 2008/ Accepted YYY 2008}

   \titlerunning{Evolution of spiral galaxies in modified gravity:  II-  Gas dynamics}

   \authorrunning{Tiret \& Combes}

\abstract{The stability of spiral galaxies is compared in modified Newtonian Dynamics (MOND) and Newtonian dynamics with
dark matter. We extend our previous simulations that involved pure stellar discs without gas,
to deal with the effects of gas dissipation and star formation. We also vary the interpolating $\mu$-function
between the MOND and Newtonian regime. Bar formation is studied and compared in both
dynamics, from initial conditions identical in visible component morphology and kinematics
(same density profile, rotation curve, and velocity dispersion). 
One first result is that the MOND galaxy evolution is not affected by the choice of the  $\mu$-function,
it develops bars with the same frequency and strength. The choice of the  $\mu$-function significantly changes the equivalent Newton models, in changing the dark matter to visible mass ratio 
and, therefore, changing the stability. The introduction of gas shortens the timescale for bar
formation in the Newton model, but is not significantly shortened in the MOND model, since it was already small. As a consequence, with gas, the MOND and DM bar frequency histograms are now more similar than without gas. The thickening of the plane occurs through vertical resonance with the bar and peanut formation, and even more
quickly with gas. Since the mass gets more concentrated with gas, the radius
of the peanut is smaller, and the appearance of the pseudo-bulge is more boxy.
 The bar strength difference is moderated  by saturation, and feedback effects, like
the bar weakening or destruction by gas inflow due to gravity torques.
 Averaged over a series of models representing the Hubble sequence, the MOND models
have still more bars, and stronger bars, than the equivalent Newton models, better fitting the observations. Gas inflows driven by bars produce accumulations at rings or pseudo-rings
at Lindblad resonances, and MOND models can reproduce observed morphologies quite well,
as was found before in the Newtonian dynamics.
}
\maketitle


\section{Introduction} 

  Galaxy dynamics and evolution depend strongly on the amount of dark matter
 assumed, on its nature, shape and radial distribution 
(e.g., Dubinski et al 1996; Debattista \& Sellwood 2000; Athanassoula 2002; 
Valenzuela \& Klypin 2003; Avila-Reese et al. 2005; Col\'{\i}n et al. 2006).
They are thus expected to be quite different under the hypothesis
of modified gravity, built to be free of any dark matter. Simulations of galaxy dynamics compared
to the observations could then provide constraints with the models.

Galaxy stability was investigated in modified Newtonian Dynamics (MOND) by Brada \& Milgrom (1999), through simulations
of dissipationless stellar discs. While it is known from Ostriker \& Peebles (1973) that dark
matter haloes have a stabilising influence on disc dynamics, they show that MOND discs
without any dark matter are also more stable than their analogs  in Newtonian gravity.
 In other words, they compare the stabilising effects of the MOND dynamics with those
of dark matter haloes, and conclude that they are both efficient, for the same disc ``temperature'',
or the fraction of the total kinetic energy in random motion. This temperature is quantified
by the ratio $t = T_{rot}/|W|$ of the rotational kinetic energy $T_{rot}$ and the absolute value of
the total gravitational energy W. If the system is stationary, the latter is equal to 
twice the total kinetic energy, through the virial theorem. The maximum value of $t$ is 0.5,
and Ostriker \& Peebles (1973) proposed a criterion for stability of $t < 0.14$.
 This criterion $t$ based on the fraction of ordered to total kinetic
energy is only one way to define the ``temperature'' of a disc to be compared in the
two dynamics.

The Toomre $Q$ parameter is often used as the criterion for axisymmetric stability
of a galaxy disc, of surface density $\Sigma$, radial velocity dispersion $\sigma_r$ and 
epicyclic frequency $\kappa$. Its value in Newtonian dynamics, which is:

\begin{equation}
\label{eq:Q_newt}
 Q = {{\sigma_r \kappa}\over{3.36 G \Sigma}}
\end{equation} 

has to be modified in MOND, by replacing $G$ by $G/\mu^+ (1 + L^+)^{1/2}$,
where $ \mu^+ $ is the value of the MOND interpolating function just above the disc, and 
$L = d ln \mu(x) /d ln x$  (Milgrom 1989). The function $\mu(x)$  satisfies
$\mu(x <<1) \sim x$ and $\mu(x >>1) \sim 1$.

The two criteria $t$ and $Q$ are therefore not equivalent in MOND, since they depend
on the total mass (Brada \& Milgrom 1999). If the  criterion of $t < 0.14$ corresponds
to about $Q > 2.5$ for Newtonian galaxy models, or MOND models of high mass, 
they correspond to much higher $Q$ for dwarf galaxies, deep in the MOND regime.
This explains the conclusion by Brada \& Milgrom (1999) that discs are 
locally more stable in the deep MOND regime.
This conclusion, however, pertains to the {\it local} stability, based on the modified
Toomre criterion. For the {\it global} stability,  Brada \& Milgrom (1999) find 
that the MOND galaxies have a larger growth rate than the Newtonian equivalent. 
Physically, the extra stability in the MOND regime comes from the fact that the gravity force
is proportional to the square of the density, instead of the density. However, the fact that
the disc is fully self-gravitating and not embedded in a stabilising halo component acts in
the other direction. 

The two proposed ``temperature'' parameters are not easy to compare with observations.
The modified Toomre parameter is model dependent, and for the first case,
 the ordered to total kinetic energy ratio $t$, the total energy involves
the random motions of the dark matter halo in the Newtonian case, which is not observed.
  To compare MOND and Newtonian dynamics with DM, we prefer to adopt the 
same properties for the visible component, both in morphology and kinematics. In addition to
identical morphology, the initial discs have the same 
rotation curve and the same radial velocity dispersion. This ``temperature''
parameter is closer to observations, although the two models
do not always have the same ``temperature'' with all possible definitions.
In a previous work, considering only stellar dynamics, we concluded that galaxies were more unstable in MOND dynamics (Tiret \& Combes 2007,
hereafter TC07).  
 The overall conclusion, averaged over a series of models representing the Hubble sequence, was that the
bar frequency is higher in the MOND dynamics, and more compatible with observations.

In the present work, we consider in addition the gas dynamics, which has a major effect on
disc stability.  We also adopt a different interpolating function $\mu$, which is the standard
one proposed by Milgrom (1983) for galaxy rotation curves. This one reduces the amount of 
``phantom'' dark matter, i.e. the dark matter required in Newtonian dynamics to obtain the same rotation
curve as in the MOND disc. In our previous work (TC07) we adopted the
$\mu(x)= x/(1+x)$ function, proposed by Zhao \& Famaey (2006),
for which the ``phantom'' dark matter is almost doubled. Therefore, in this paper, we recompute all models,
with and without gas, to provide better comparisons.

In the next section, we describe the numerical techniques used to solve the modified Poisson
equation, and also the model of sticky particles adopted for gas dynamics. Galaxy models and initial
conditions are displayed in Sect. 3, results are presented in Sect. 4 and discussed in Sect. 5,
to establish the new bar frequency in the MOND and Newtonian models.

\section{Numerical model}

\paragraph{Gravitation.}

We use our potential solver described in TC07, to solve the Poisson equation in the case of 
Newton gravity (Eq. \ref{eq:poison_N}) and MOND (Eq. \ref{eq:poison_MOND}):

\begin{eqnarray}
\label{eq:poison_N}
   \bigtriangleup  \phi&=&4\pi G \rho \\
\label{eq:poison_MOND}
  \overrightarrow{\bigtriangledown} \cdot \left[ \mu\left( \frac{\mid \overrightarrow{\bigtriangledown} \phi \mid}{a_0} \right) \overrightarrow{\bigtriangledown} \phi \right]&=&4\pi G \rho
\end{eqnarray} 

\noindent This grid solver uses multigrid techniques (Numerical Recipes, Press et al., 1992), the grid characteristics are the same as in TC07. The simulation box covers a $100\ \mathrm{kpc}$ cube, the spatial resolution is about $400\ \mathrm{pc}$ ($256^3$ nodes).

\paragraph{Gas dissipation.}

These new simulations include a gas component. The dissipation between \textit{gas clouds} is modelled 
by a sticky-particle scheme. This allows us to control more directly the gas dissipation,
through a rebound, or elasticity parameter $\beta$ (e.g., Combes \& Gerin 1985).

Consider two particles ($\vec r_1$, $\vec v_1$, $m_1$), ($\vec r_2$, $\vec v_2$, $m_2$) in their 
mass centre frame. The radial and tangential relative velocities are ($\vec v_r$, $\vec v_t$) before the collision, and ($\vec v'_r$, $\vec v'_t$) after the collision:
\begin{eqnarray*}
\vec v'_r &=& \beta_r \vec v_r \\
\vec v'_t &=& \beta_t \vec v_t
\end{eqnarray*} 
or,
\begin{eqnarray*}
\vec v'_1 &=& \frac{m_2}{M} \left[  \beta_t(\vec v_1-\vec v_2) + (\beta_r - \beta_t) ( (\vec v_1-\vec v_2)\cdot \vec e_r )\vec e_r  \right] \\
\vec v'_2 &=& \frac{m_1}{M} \left[  \beta_t(\vec v_2-\vec v_1) + (\beta_r - \beta_t) ( (\vec v_2-\vec v_1)\cdot \vec e_r )\vec e_r  \right]
\end{eqnarray*} 
with $M=m_1+m_2$ and $\vec e_r = (\vec r_2-\vec r_1)/\parallel (\vec r_2-\vec r_1) \parallel$.

We choose $\beta_t = 1$ to conserve the angular momentum, and $\beta_r = -0.85$. $\beta_r<0$ 
ensures that two particles after the collision go in opposite ways than before the collision. This also introduces a randomisation in velocities.

Selecting $\beta_r>0$ results in a model with colder gas, and lower velocity dispersion.

The collision grid, used to find the neighbours for collisions, is in 2D due to the small thickness of the gas layer. 
The collision cell size is the same as the gravitational one so that dissipation cools at the same resolution scale 
than gravity heats. 
An individual gas cloud is typically subject to $10-20$ collisions along a galactic rotation.

\paragraph{Star formation.}

The star formation follows the Schmidt law:
$$ {{dN_\star}\over{dt}}=C\cdot \rho_{gas}^{1.2} $$
where C is calibrated so that gas consumption time through
star formation is $\tau\sim 2$ Gyr for a spiral galaxy.
We use the hybrid particle technique to form stars from gas. Since the amount of stellar material 
formed at a given location is always smaller than the particle mass resolution, only a fraction of the gas particle
is transformed into stars. The rest remaining gaseous, until several cycles use up the total gas content of
a particle. For the dissipation rate, in the sticky-particle scheme, the effective mass fraction
 of gas is taken into account. When the gas fraction of a given particle is less than 
$0.5$, $\beta_r$ is chosen to be positive 
(the particles cross each other and do not rebound during the collision). As the gas fraction decreases, the 
dissipation is progressively reduced, and the hybrid particle takes on a stellar behaviour.

\section{Initial conditions}

\begin{table}

\centering
\begin{tabular}{c c c c c c c c}
\hline\hline 
 Run & $M_d$ & $M_b$ & Gas   &$a_d$ & $a_g$ & $M_h$& $b_h$ \\
\hline
Sa   & 40    & 12.7  & $2\%$ & 4    & 6   & 146.3  & 29.4   \\
Sb   & 30    & 5     & $5\%$ & 5    & 7.5 & 118.9  & 24.8   \\
Sc   & 20    & 2     & $7\%$ & 6    & 9   & 92.8   & 20.9   \\
\hline
\end{tabular} 
\caption{Parameters of the Sa, Sb, Sc galaxy models. The unit system is $G=1$: the mass unity is $U_m=2.26\ 10^9\ M_\odot$, the length unity is $U_r=1.02\ kpc$.
}
\label{tab:param}
\end{table}

\subsection{The $\mu$-function}

In TC07, we used the \textit{simple} form of the $\mu-$function: $\mu(x)=x/(1+x)$, proposed by Zhao \& Famaey (2006).
This function was to be a better compromise with theory and observations of the Milky Way than 
other functions like the 
\textit{standard} one: $\mu(x)=x/\sqrt{1+x^2}$. However, most galaxy rotation curves
are reproduced with the standard function (e.g, Sanders \& McGaugh 2002). 
Coupled with $a_0=1.2\times10^{-10}\ \mathrm{m.s^{-2}}$, 
the well-known value for the critical acceleration of MOND, the rotation curve of the galaxy is thus fixed if one gives 
a density profile. In our study, we will always compare any MOND galaxy model with the Newtonian gravity \& dark 
matter model (DM) having the same rotation curve. 
In this paper, we use the \textit{standard} $\mu-$function.

\subsection{The galaxy models}

We study in the present paper,
high-surface brightness galaxies, while low-surface brightness (LSB) galaxies will be modelled in future works.
LSB systems require a full study, since bar formation is qualitatively
different there (e.g., Mayer \& Wadsley 2004). 

As in TC07,  the stellar disc is modelled by a Miyamoto-Nagai disc. 
The initial distribution of the gas follows a Toomre disc:
$$\Sigma_g(r)={{a_g M_g}\over{2\pi (r^2+a_g^2)^{3/2}}}\ .$$
The 3D structure is given by:
$$\rho_g(r,z)=\Sigma_g(r) \mathrm{sech}^2(z/H_g),$$
$a_g$ and $H_g$ being the radial and vertical gas scale-lengths. 
The bulge and the DM halo are initially distributed in Plummer spheres live haloes.
Compared to TC07 the stellar matter (disc+bulge) keeps the same characteristics. Now we add 
a gas disc of $2\%$ to $7\%$ of the total visible mass. 
The mass and the characteristic radius of the ``phantom'' dark halo (in the corresponding Newtonian dynamics)
have changed because we now use the standard $\mu$-function 
instead of the simple $\mu$-function. Simulations without gas have been repeated with these new parameters for the DM and MOND models.
The characteristic values are given in the Table \ref{tab:param}. The characteristic length of the bulge for Sa, Sb, Sc 
galaxies is $1$ kpc. For the characteristic height of the Miyamoto-Nagai disc, we choose $b_d/a_d=1/10$. The given 
mass is the truncated mass. The Toomre parameter value is the same in the DM and MOND model: $Q=2$. 
We use the same definition of Q for the Newtonian and MOND simulations:  
$Q = {{\sigma_r \kappa}\over{3.36 G \Sigma}}$, where G is not varied.  Instead of trying to adapt
the stability status (cf Introduction section), we  build
the initial galaxy to have the same morphological and kinematical 
characteristics in both models, in what concerns the visible component.
 The 
number of particle is $2.10^5$ for the stellar disc; $5.10^4$ for the Sa/Sb gas discs; and $10^5$ for the Sc gas disc. 
The mass of a bulge particle is equal to the mass of a disc particle. The mass of dark matter particles is three times 
the mass of the disc particle. The initial distribution of star, gas, and dark matter particles are respectively truncated at 
$25$ kpc, $30$ kpc, and $45$ kpc.

\section{Results}

\subsection{The couple of parameters ($\mu, a_0$)}

\begin{figure}
  \resizebox{\hsize}{!}{\includegraphics{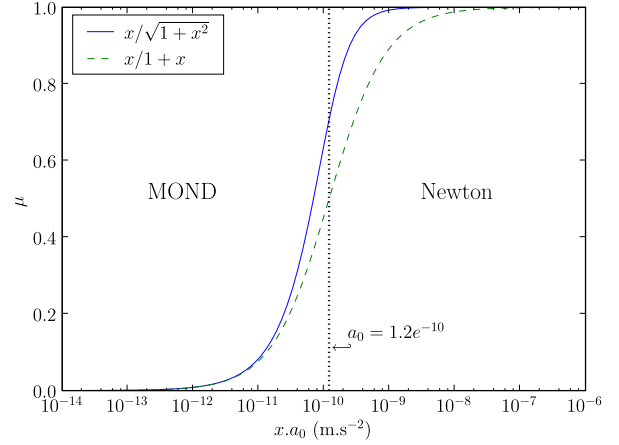}}
  \caption{The simple $\mu$-function ($\mu(x)=x/(1+x)$, used by TC07) transfers in the MOND regime one order of magnitude  
earlier in acceleration, than the standard $\mu$-function ($\mu(x)=x/\sqrt{1+x^2}$), used in the present work. }
  \label{fig:muplot}
\end{figure}

\begin{figure}
  \resizebox{\hsize}{!}{\includegraphics{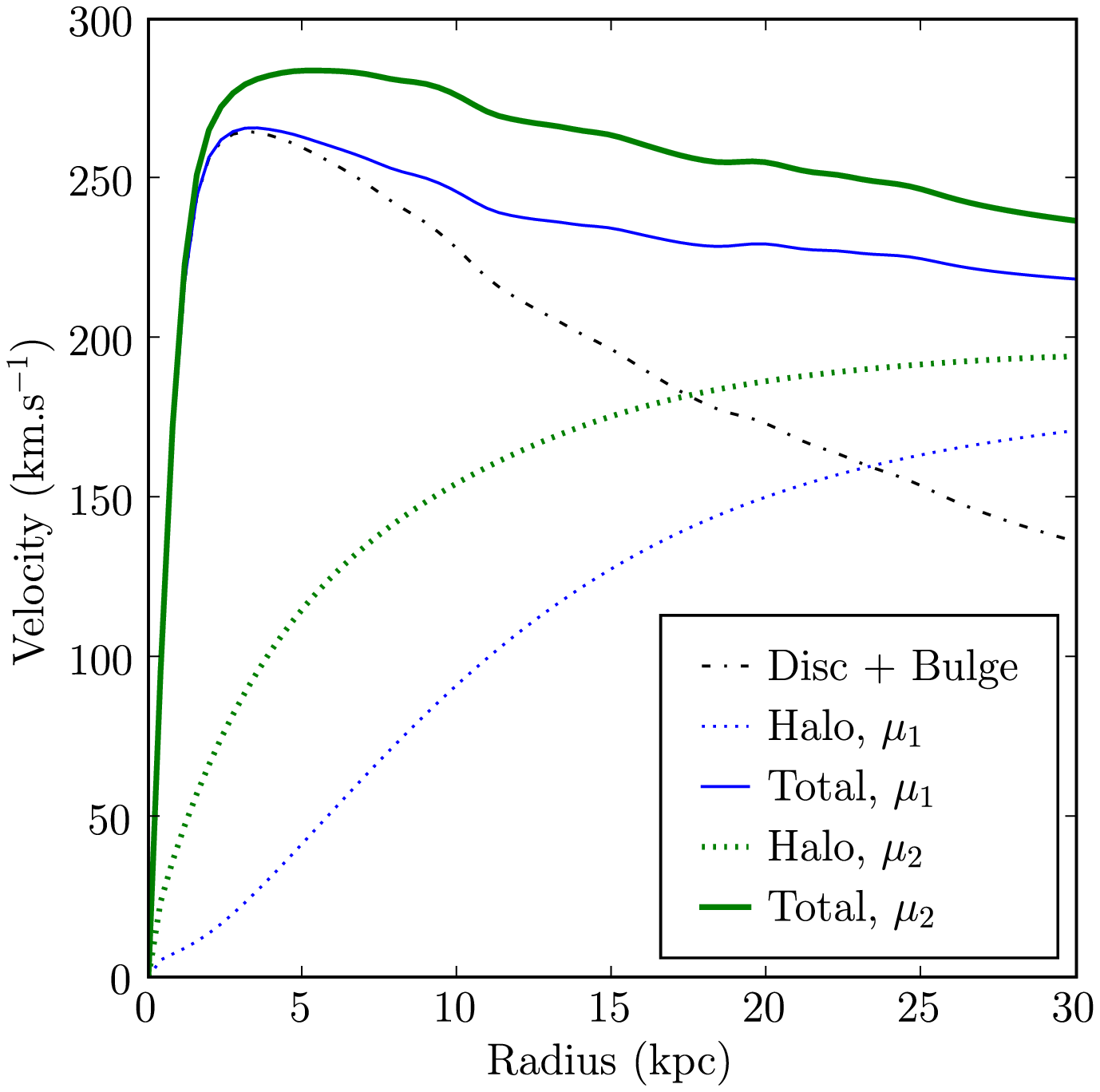}\includegraphics{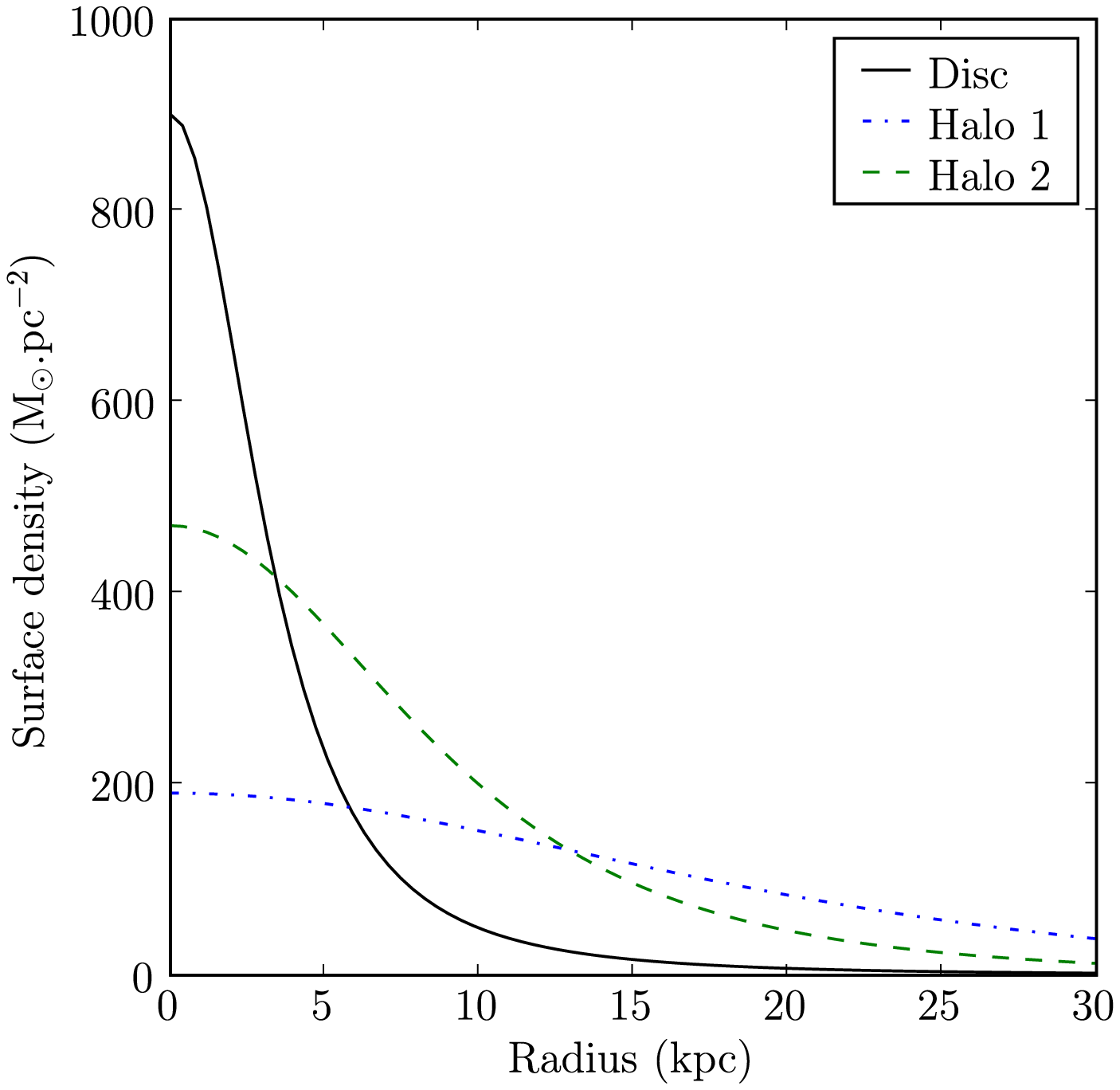}}
  \caption{Left: rotation curves of an Sa-type galaxy using the \textit{standard} ($\mu_1$, this work) and \textit{simple} ($\mu_2$, TC07) 
$\mu$-functions.  In each case, the halo needed in the DM model to obtain the MOND rotation curve is plotted too (Halo,  $\mu_1$, Halo, $\mu_2$ 
curves). Right: Surface density of the stellar disc compared to the dark matter halo (in the Newtonian model) needed to fit the MOND
 rotation curve if the free function is the standard one (1, this work) or the simple one (2, TC07).}
  \label{fig:rc}
\end{figure}

In TC07, we found that the bar instability occurs in a much longer timescale in the DM model than in MOND (several Gyr 
compared to less than one). By adding gas, the bar instability could be expected to develop faster (about $1\ \mathrm{Gyr}$) 
especially in the DM model. First, we run several simulations, keeping the ``simple'' $\mu$-function; 
the bar appears sooner with gas but not as soon as could have been expected. The dark matter halo needed to fit the MOND 
rotation curve (using the simple $\mu$-function) is very massive and stabilizes the galactic disc too much.

With the simple 
$\mu$-function, the acceleration stays Newtonian until $a=10^{-8}$ m.s$^{-2}$ (Fig. \ref{fig:muplot}). This means that below $a=10^{-9}$ m.s$^{-2}$ in the DM-model, the gravitation begins to be dominated by the dark matter halo.

To increase the Newtonian regime, and avoid introducing too much dark matter for the DM model, there are two 
possibilities: either change the value of the critical acceleration $a_0$, or change the standard function and keep 
$a_0=1.2\times10^{-10}\ \mathrm{m.s^{-2}}$.
 The latter possibility is adopted in the present work. 
Figure \ref{fig:rc} (left) shows an example of rotation curve computed with the simple and standard 
function. At first sight, the difference between the two curves does not seem very important (about $25\ \mathrm{km.s^{-1}}$), 
but the dark matter mass needed to fit the MOND rotation curve with the standard $\mu$- function 
is twice the dark matter mass 
necessary for the simple $\mu$-function (Fig. \ref{fig:rc}, right), inside the visible radius. 
This plot corresponds to an Sa-type galaxy, where
the more massive and concentrated dark matter (halo 2) multiply by a factor 2 the time to form the bar (see TC07). 
It is even more sensitive for the late-type galaxies which have an even larger fraction of dark matter. 
We do not think that this stabilising effect is due to 
a different swing amplification: with the first
developing pattern, we computed the  
parameter $X=\lambda/\sin i/\lambda_{crit}$ where $i$ is the pitch angle of the 
spiral arms, and $\lambda_{crit}=4\pi^2G\mu/\kappa^2$. The value for the halo 1 and halo 2 models are 
comparable: $X\sim 1$.

\begin{figure*}[!]
\resizebox{\hsize}{!}{\includegraphics{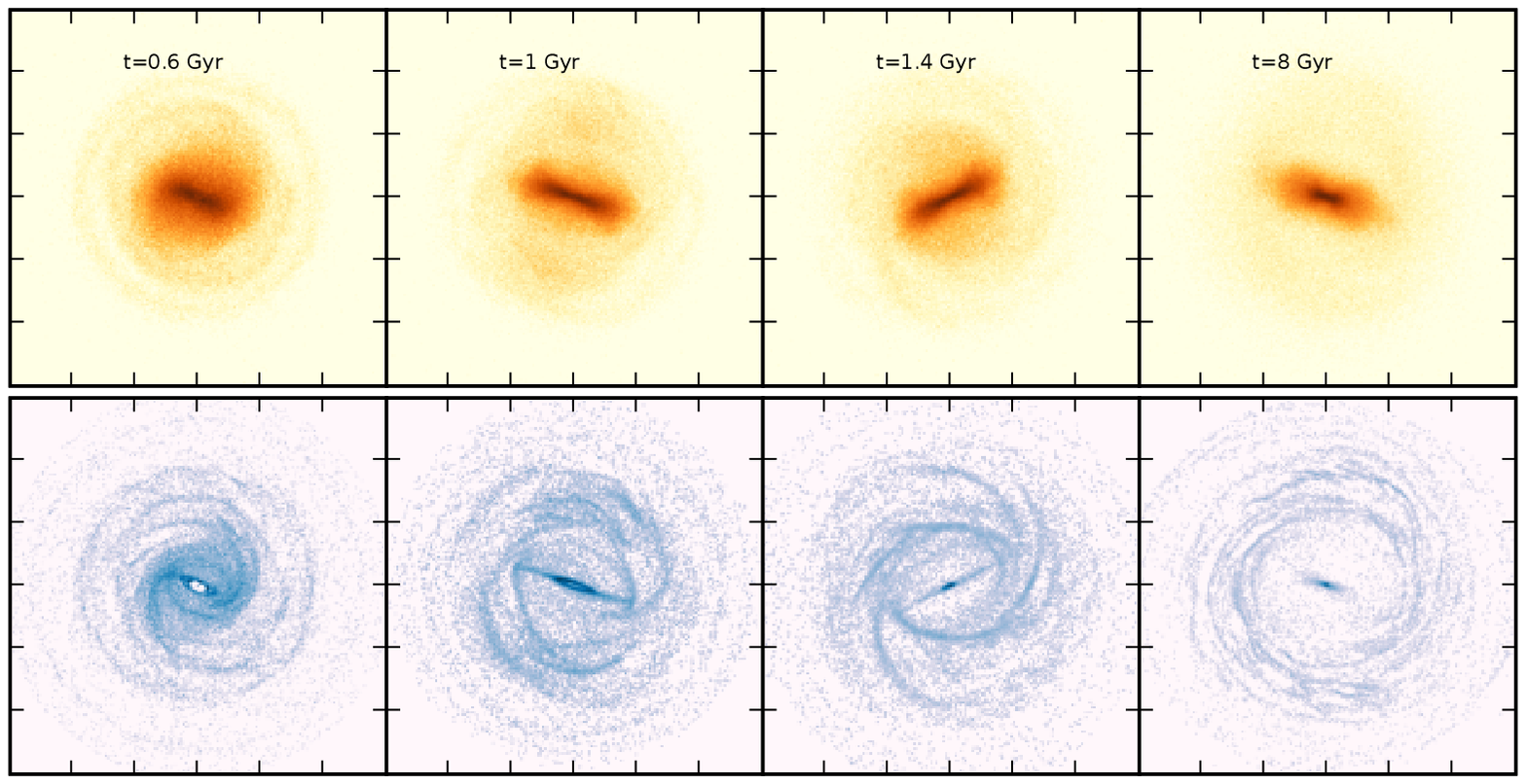}}
\resizebox{\hsize}{!}{\includegraphics{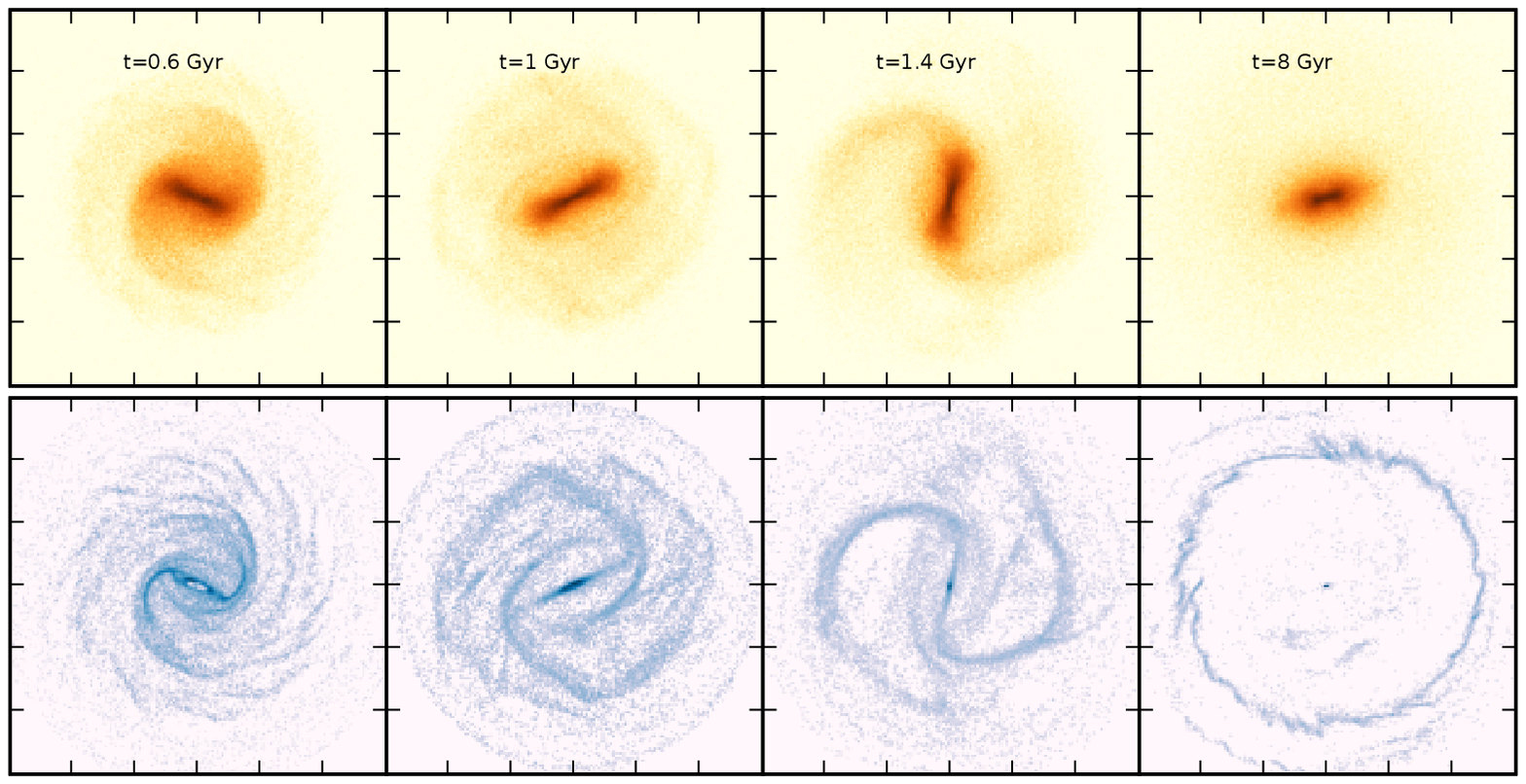}}
     \caption{Snapshots of an Sa type galaxy in the DM model (top) and in the MOND model (bottom).  In each
row, the stellar component is shown above, and the gas below, for times going from 0.6 to 8 Gyr. These evolutions show in DM and in MOND the bar instability, with the formation of grand design spiral arms, and the successive apparitions of ILR, UHR, and OLR rings (see text). The size of the box is $80$ kpc$\times 80$ kpc.}
     \label{fig:sa_dm}
\end{figure*}

\subsection{Bar strength}

\begin{figure}[!]
  \resizebox{\hsize}{!}{\includegraphics{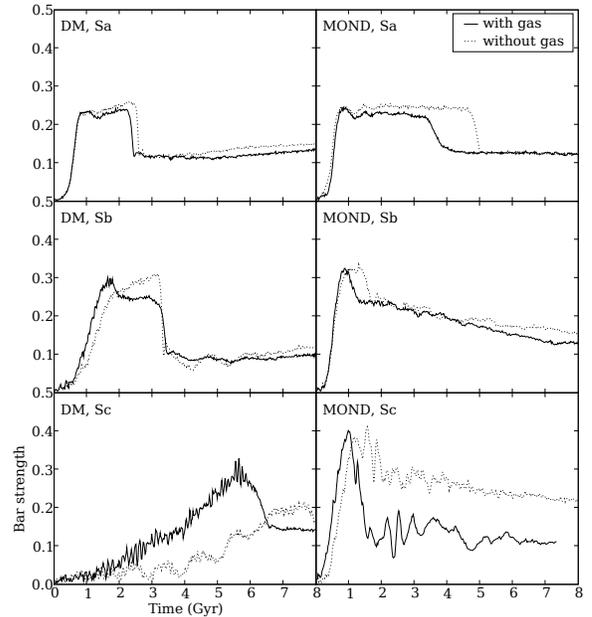}}
  \caption{Several cases of bar strength evolution in DM and MOND models, with and without gas. Bars are formed in $1-2$ Gyr in all models, 
except for the Sc galaxy in DM, where the halo is relatively more dominant and stabilises the disc.}
  \label{fig:barstrength}
\end{figure}

By losing energy during collision (dissipation), the gas remains a cold component that tends to destabilize the galaxy disc. With gas, 
bar formation is thus expected to be faster. In gas-poor, early-type galaxies (Sa, Sb), the gas contribution to the galaxy evolution is weak 
(Fig. \ref{fig:barstrength}).
The difference appears more clearly for the Sc galaxy (with $7 \%$ of gas). 
With gas, the bar forms sooner and is stronger. The bar strength is estimated by the quantity 
$Q_2= Max\vert F_{\theta}(r)/F_r(r)\vert $, where $F_r$ is the radial force, and $F_{\theta}$ the tangential force, corresponding 
to the Fourier component $m=2$. In the DM model, it is maximum at $t = 6\ \mathrm{Gyr}$ ($Q_2 = 0.3$) without gas instead of 
$t = 7.5\ \mathrm{Gyr}$ ($Q_2 = 0.2$) with gas. In the Sa, Sb simulations, the bar is weakened by the apparition of a peanut resonance 
(see TC07). This occurs sooner when there is gas. Quite rapidly, the gas falls in the centre of the galaxy, increasing the potential well 
through its mass concentration. We observe that galaxies in MOND spend more time with a stronger bar than in the DM model. In 
the MOND-Sa simulation, the peanut that weakened the bar occurs later ($t=~4$ Gyr) than in the DM-Sa simulation ($t=~2$ Gyr). 
In the DM-Sb simulation, the bar is strongly weakened after the vertical buckling: $Q_2=0.1$, while it stays at $Q_2=0.2$ during 
several gigayears after the resonance in the MOND simulation.

\begin{figure}[!]
  \resizebox{\hsize}{!}{\includegraphics{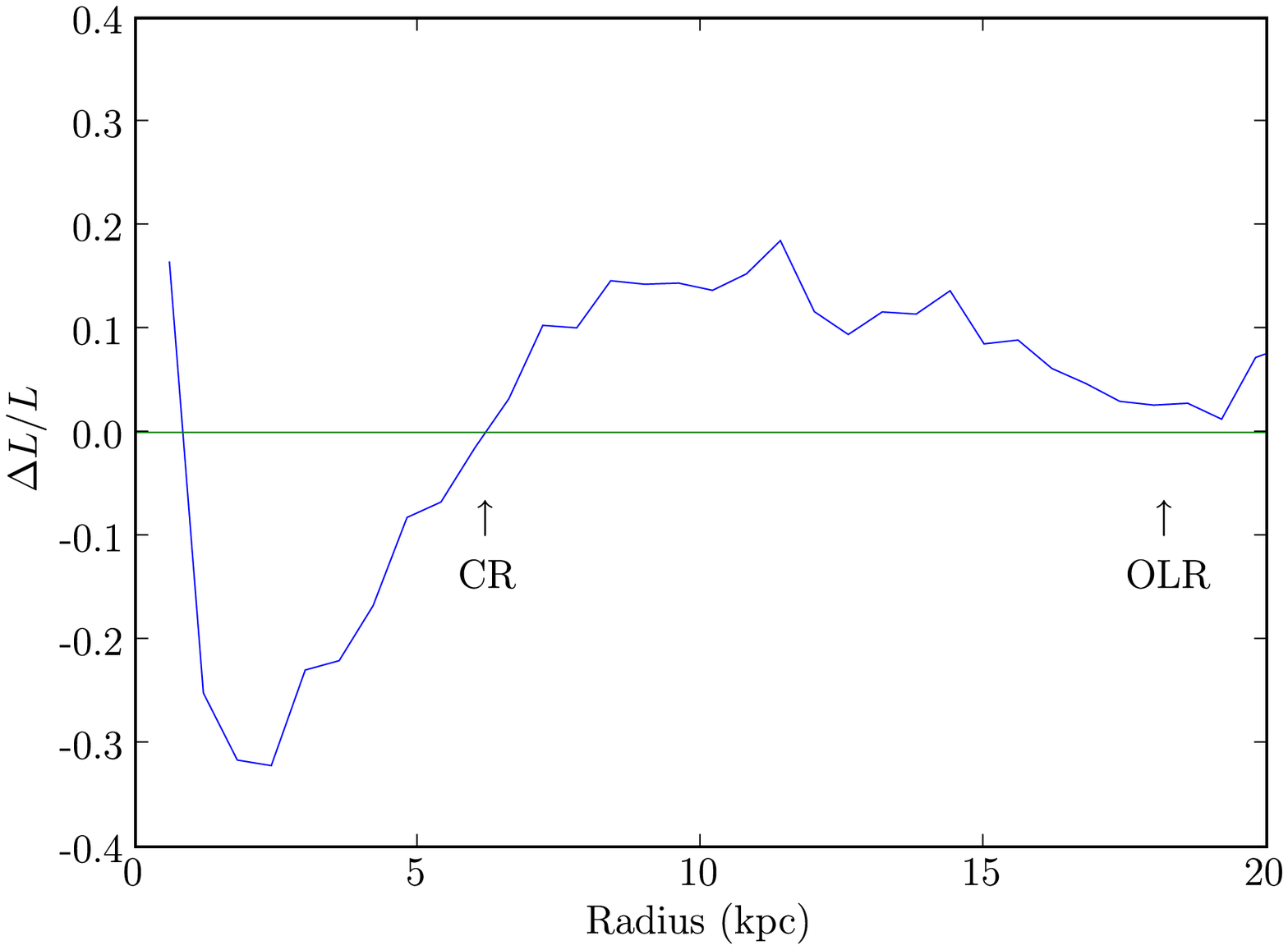}}
  \caption{Effects of the gravity torques created by the stellar bar during the bar dissolution (MOND, Sc: $t\sim 1.2$Gyr). 
It correspond to the variation of angular momentum over one galactic rotation normalized to $L(r)$. The gas loses 15-30\% 
in region of the bar during a short period of $100$ Myr.  }
  \label{fig:torque}
\end{figure}

The MOND-Sc simulation shows that the bar is destroyed immediately after its formation because of gravity torques and
subsequent gas inflow, and angular momentum exchange, as already shown by Bournaud et al. (2005). To maintain
or reform bars, gas accretion should then be considered (Bournaud \& Combes, 2002). 
Figure \ref{fig:torque} shows the angular momentum lost by the gas along one galactic rotation.

\subsection{Bar pattern speed and Resonance}

\begin{figure}
  \resizebox{\hsize}{!}{\includegraphics{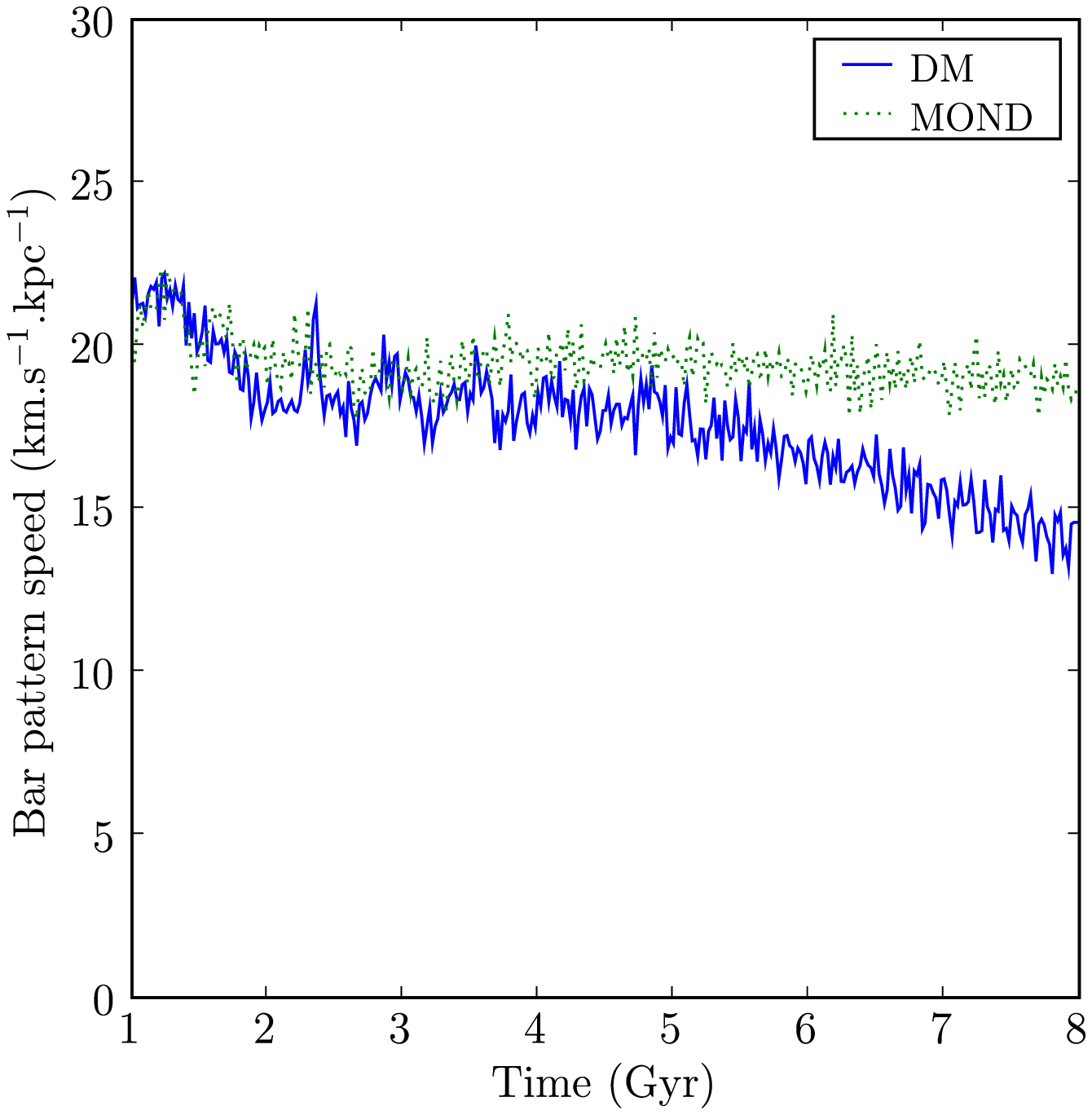}\includegraphics{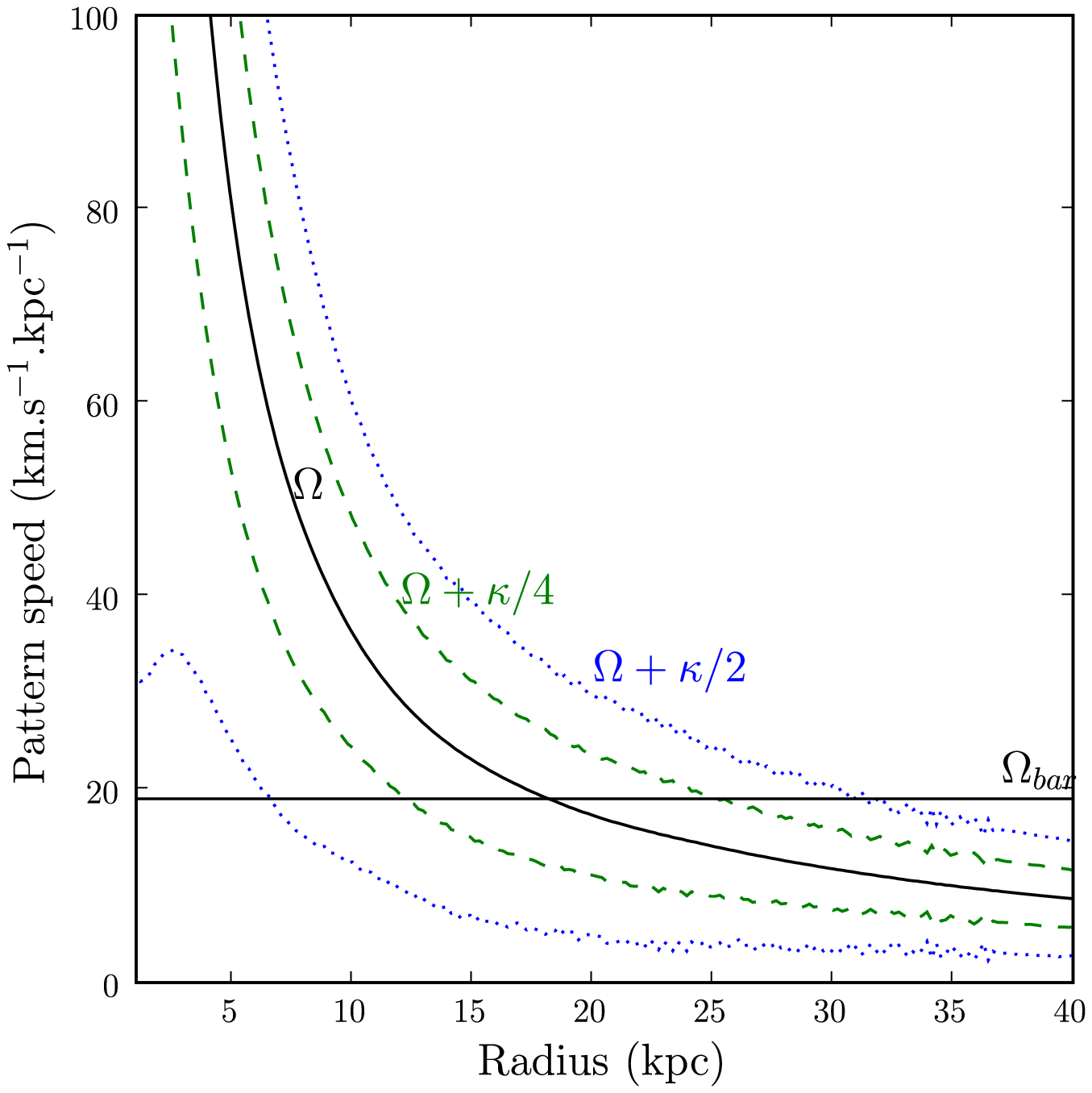}}
  \caption{Left: in the DM model, the stellar bar is submitted to the dynamical friction against the dark matter
particles in the halo; the bar pattern speed slows 
down in the DM model and keeps constant in the MOND model. Right: Pattern speed and location of the Lindblad resonances in the MOND model at $t=8\ \mathrm{Gyr}$.}
  \label{fig:dfric}
\end{figure}

Many authors have studied the dynamical friction between the galactic bar and the dark matter halo, 
and the subsequent bar slow down and angular momentum transfer (Weinberg 1985; Little \& Carlberg 1991; 
Hernquist \& Weinberg 1992; Debattista \& Sellwood 1998 2000; O'Neill \& Dubinski 2003; 
Valenzuela \& Klypin 2003; Sellwood \& Debattista 2006). As in TC07, we make a comparison between the DM and MOND models about the bar pattern speed and its implication on the resonances.

The bar pattern speed ($\Omega_{bar}$) is calculated from the Fourier transform $
\widehat{\phi_2} (r,\Omega)$ of $\phi_2(r,t)$, the phase term of the potential Fourier decomposition 
(see TC07). In the DM model, the bar pattern speed loses $5-10\ \mathrm{km.s^{-1}.kpc^{-1}}$ after its formation, 
while it stays constant 
in the MOND model (Fig. \ref{fig:dfric}, left). In TC07, we attribute this slow down to dynamical friction effects. 
The slow down is less pronounced than in the previous work because the dark matter halo is less massive and the dynamical friction 
experienced by the stellar bar against the dark matter halo is less important.

Let us come back to the dominant difference between MOND and DM models as far as dynamical 
friction is concerned. A previous analysis by Ciotti and Binney (2004) has shown that, due to the much
longer range of the MOND gravity, dynamical friction is more efficient. However, there is no dark
matter particles to transfer angular momentum, and this considerably reduces the friction for MOND. 
Recently Nipoti et al. (2008) studied the slow down of a rigid bar, built as a small perturbation (a few 
percent in mass) in an otherwise stellar disc and find that the dynamical friction is more efficient 
in MOND than in DM. But again this does not reproduce the realistic case of a massive bar in an unstable disc.
In MOND, the massive bar cannot transfer its angular momentum to halo particles, and the friction is then
much lower, even negligible with respect to the DM bar model.

Fig. \ref{fig:sa_dm} shows different snapshots of an Sa-type galaxy, evolving in the DM and MOND model. The stellar and 
gaseous components have been separated. Resonant rings are easily identified in the gas disc. During the first gigayear, we 
notice the formation of an ILR (the small ring inside the bar). In this high-density region, the gas particles collide frequently and 
lose angular momentum. 
Progressively, gas particles lose energy through collisions, and the ring migrates inward with time, 
 while particles infall toward the centre of the galaxy. This depends on many factors, such as the density 
of particles in the various radial zones, dominated by the $x1$ or $x2$ orbits, the mass concentration,
and the relative strength of gravity torques and viscous torques (e.g. Schwarz 1984; Combes \& Gerin 1985; Piner et al 1995;
Regan \& Teuben 2003).
Around $t=1-1.2$ Gyr, another ring is forming at the UHR (the elongated ring surrounding the bar). 
After $t=2$ Gyr this ring is diluted in the DM model; while in the MOND model transition spirals continue 
to appear and drag particles outward, until the OLR (at $t=8$ Gyr). Figure \ref{fig:dfric} (right)
represents the rotation curve $\Omega(r)$ and combination with the epicyclic frequency $\kappa(r)$. 
The OLR is located at $30$ kpc, which is the intersection between $\Omega_{bar}$ and $\Omega + \kappa /2$.

\begin{figure}
\centering
\begin{tabular}{|c|c|}
	\hline
	\includegraphics[width=3.1 cm]{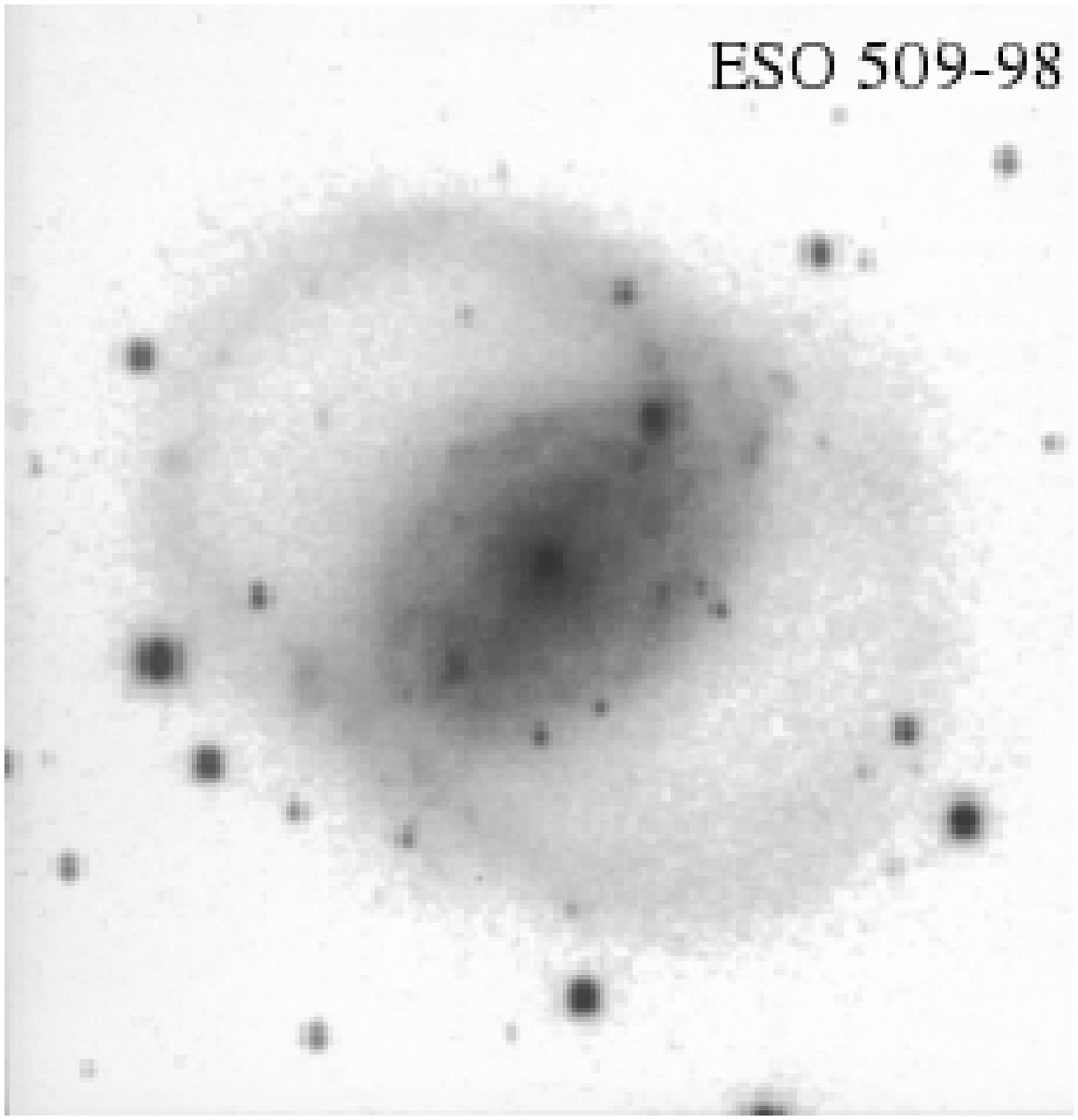}&
	\includegraphics[width=3.1 cm]{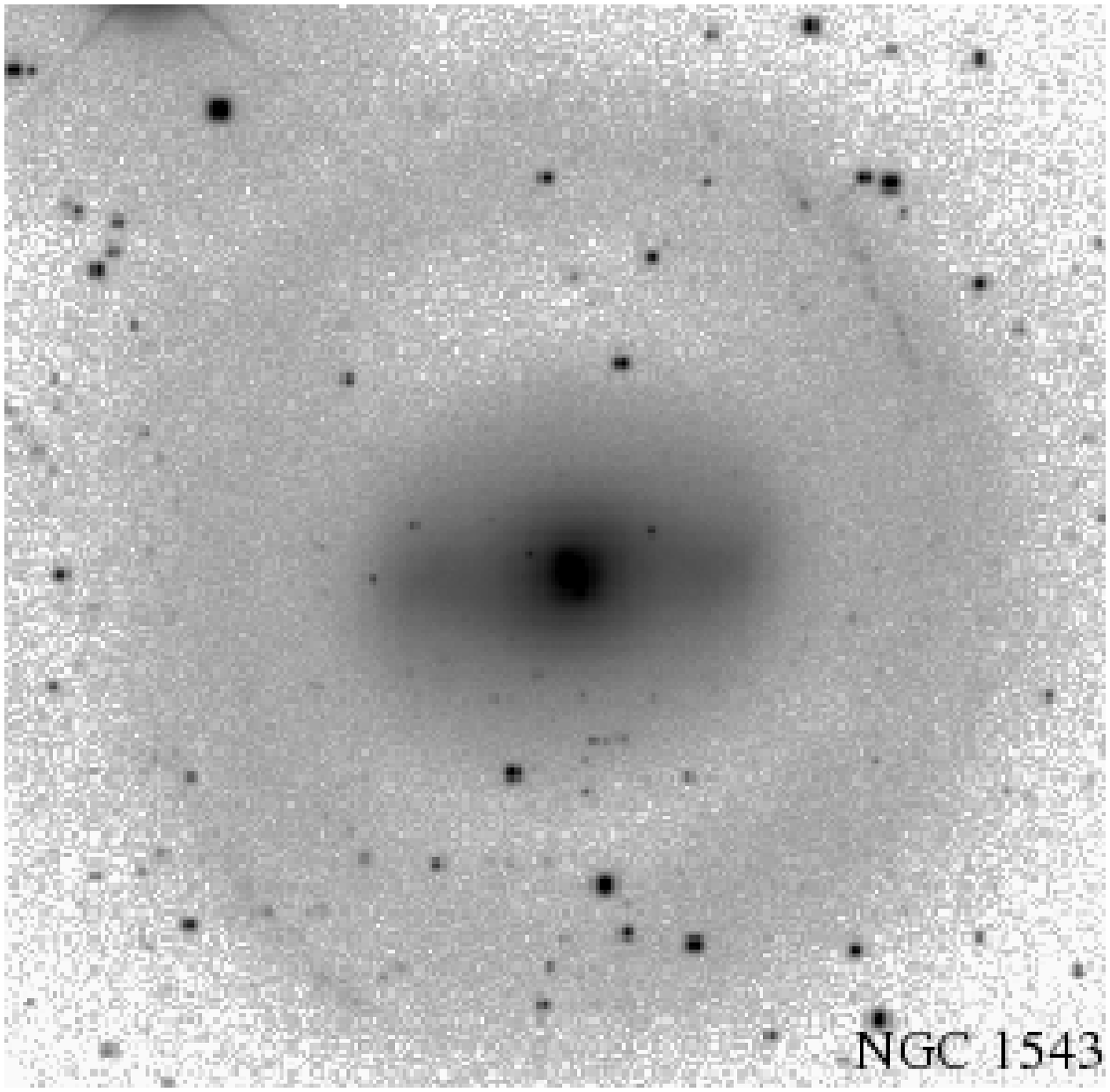}\\
	\hline
\end{tabular}
\centering

\resizebox{\hsize}{!}{\includegraphics{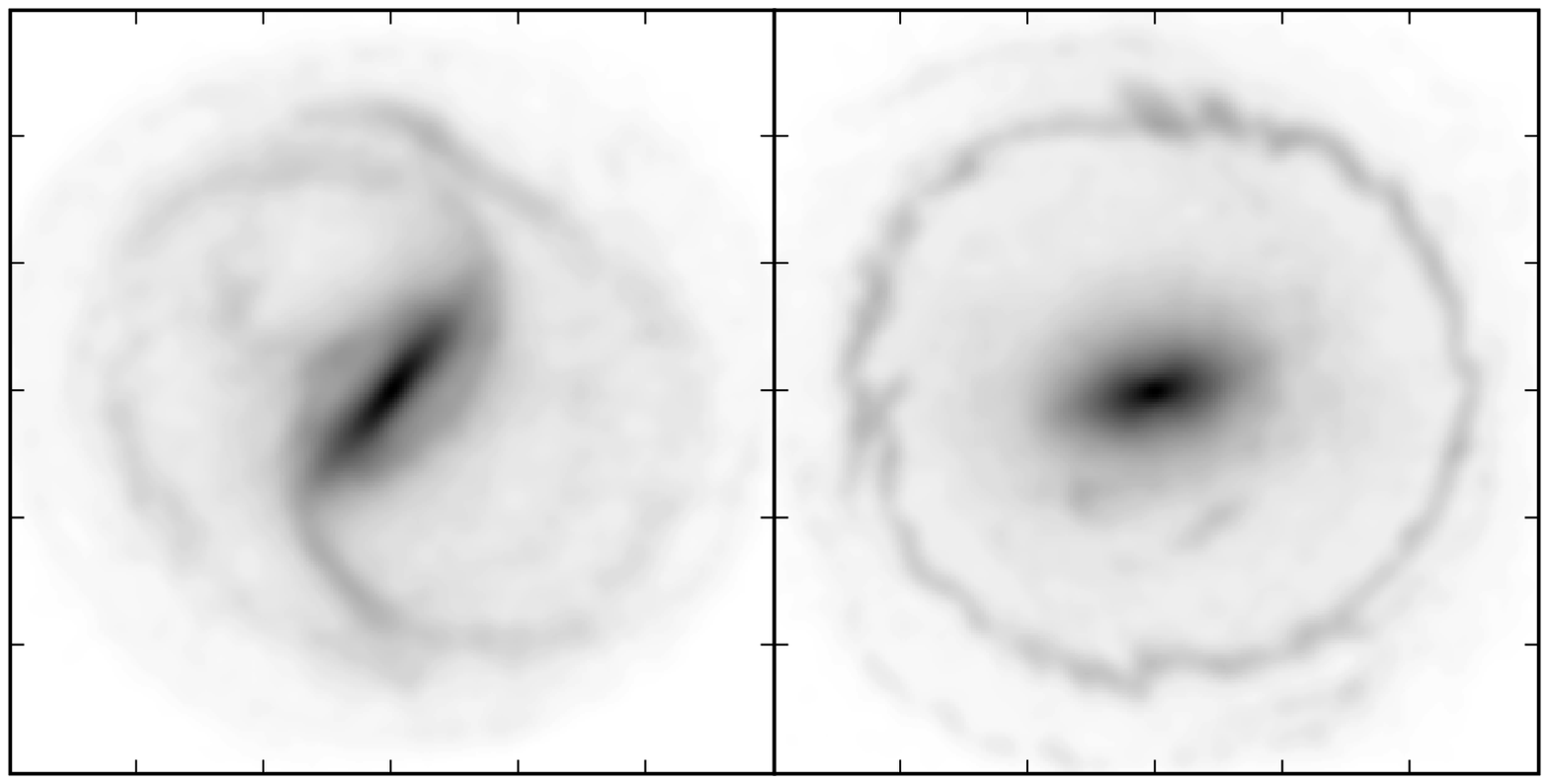}}

\caption{Two examples showing the morphological structures of ESO 509-98 and NGC 1543 
(top panel) compared to simulated galaxies in MOND (bottom panel). Rings and pseudo-rings structures are well reproduced 
with modified gravity. The size of the box in the simulation (bottom panel) is $80$ kpc$\times 80$ kpc. }
\label{fig:ring}
\end{figure}

This evolution in MOND is a good illustration of the bar formation and angular momentum transfer. 
The initial disc is not stable enough to avoid the bar instability. In the central region, particles which were on quasi-circular orbits lose angular momentum, and run on elongated orbits. The total angular momentum has to be conserved 
so it is transferred to the external region of the disc (by the spiral arms). Then, particles accumulate  either in the centre 
or in the periphery (OLR) of the galaxy.  In the simulation, the first resonance to appear is the ILR, then the UHR,
and then the OLR. In a real galaxy, it is possible to have the three simultaneously, if accretion of
external gas is taken into account, to replenish the gas disc.
The non-existence of a massive halo in MOND allows particles to settle down on stable resonant orbits. 
The bar pattern speed stays constant so the resonances remain at the same positions. Particles are trapped on these orbits more easily. 
Figure \ref{fig:ring} displays the formation of rings and pseudo-rings in MOND simulations compared to the observations.

\subsection{Vertical structure}

\begin{figure}
 \centering
 \resizebox{\hsize}{!}{\includegraphics{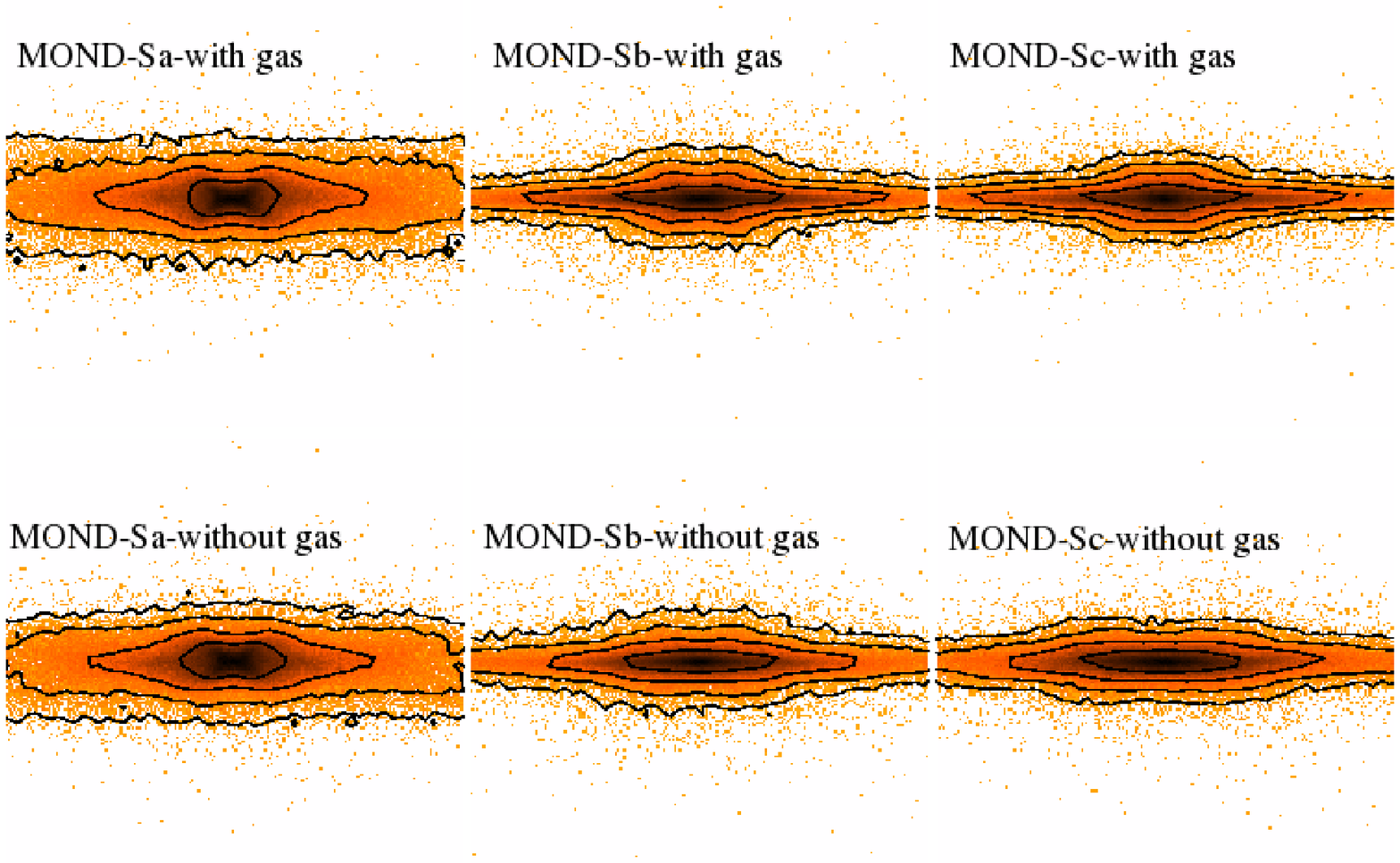}}
 \resizebox{\hsize}{!}{\includegraphics{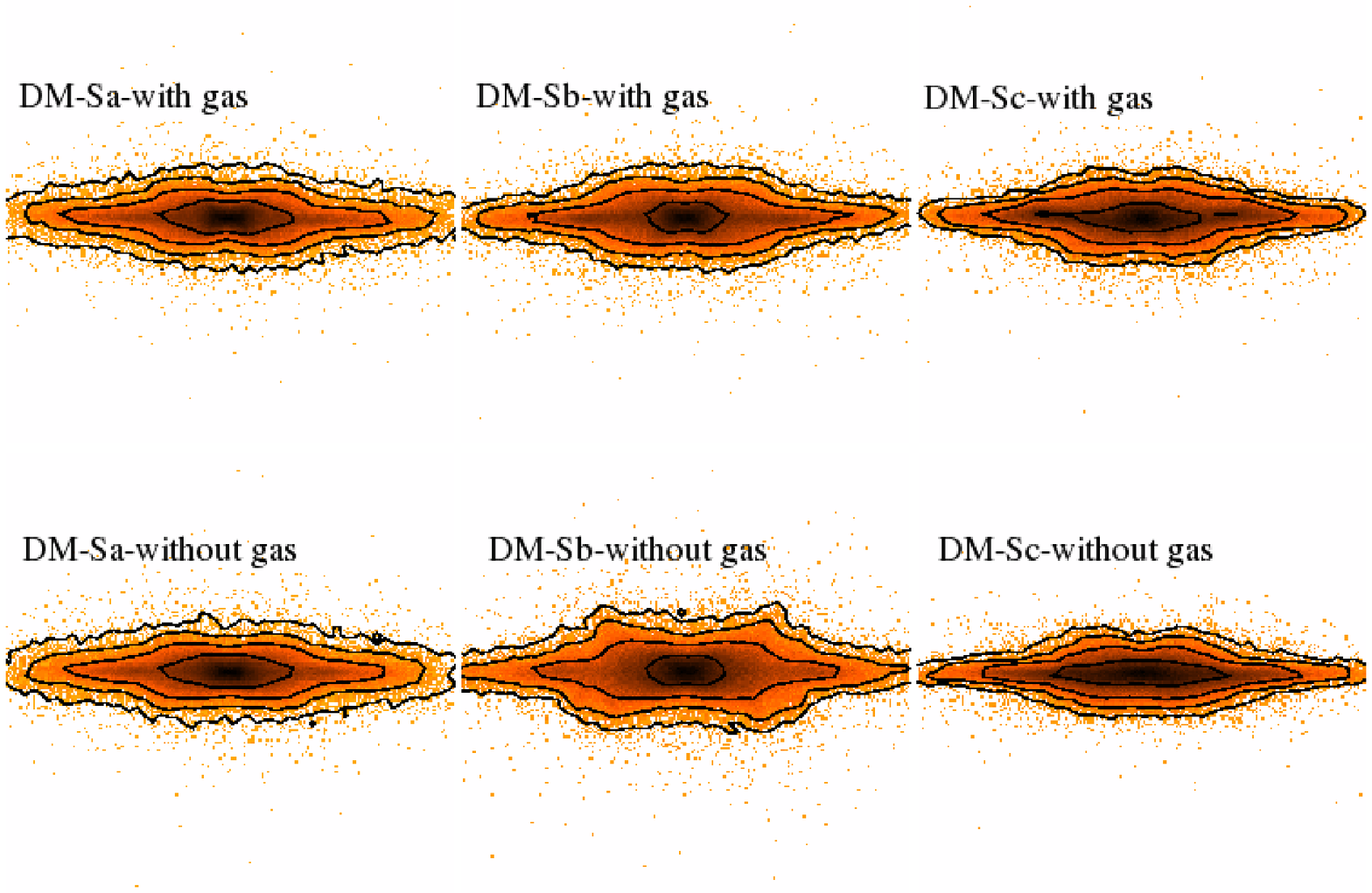}}

\caption{Edge-on views of the component of galaxies after 8 Gyr, where the bar is perpendicular to
the line of sight.
The DM simulations show a peanut structure for the Sa and Sb models, with and without gas; the Sc model begins to bend at this time. For the MOND simulations, the Sa model forms a well-defined peanut as the Sb model with gas. The Sb and Sc models without gas are rather boxy. For each galaxy type, the contour levels are the same. }

\label{fig:peanut}
\end{figure}

The edge-on views of galaxies are plotted in Fig. \ref{fig:peanut}, in the MOND and DM models, comparing the simulations with gas and without gas. Those views correspond to the final snapshot of each simulation ($t=8$ Gyr). For all the simulations with gas (in the MOND and DM models) the peanut resonance is already well settled.

\begin{figure}
 \centering
 \resizebox{\hsize}{!}{\includegraphics{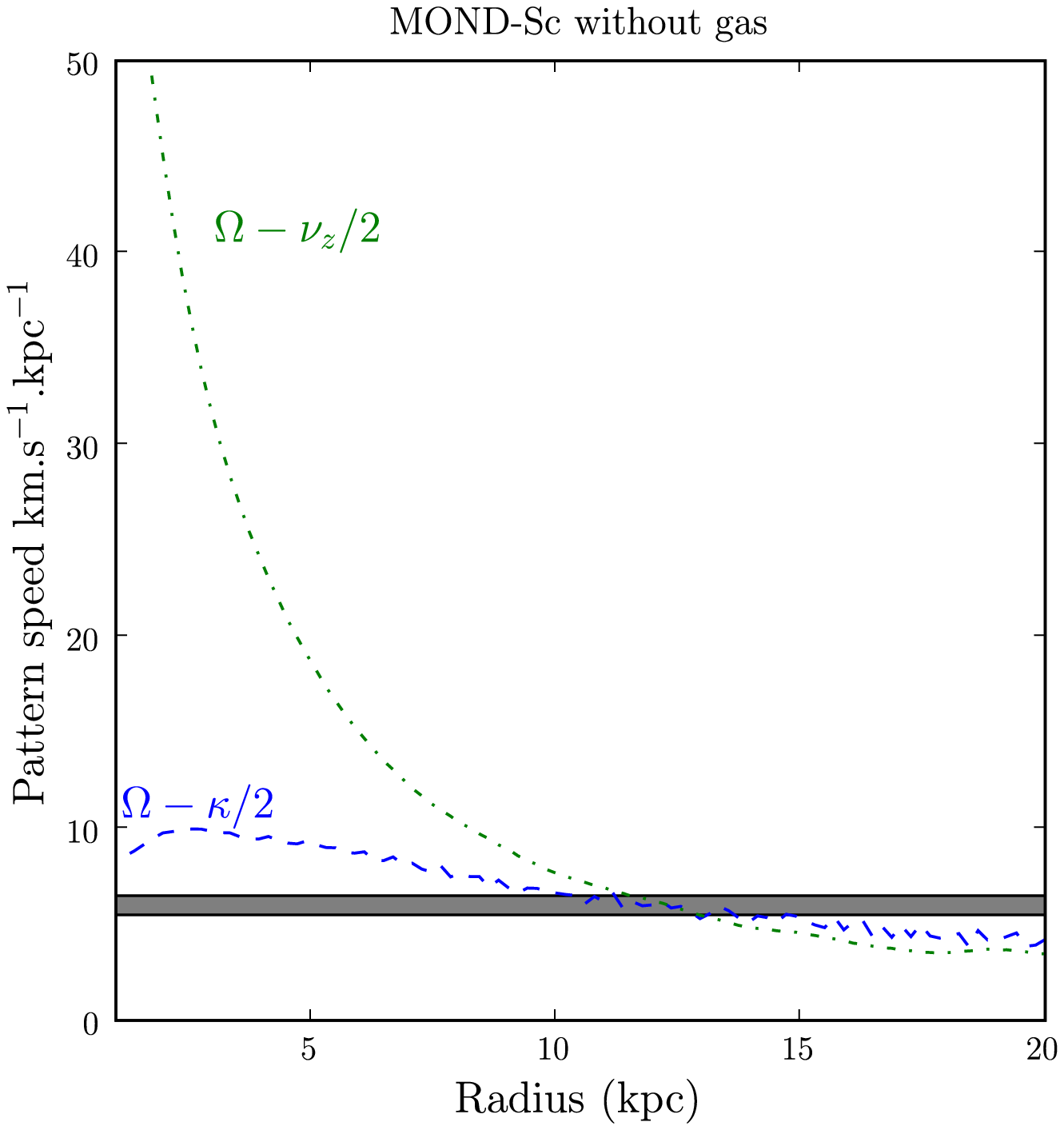}\includegraphics{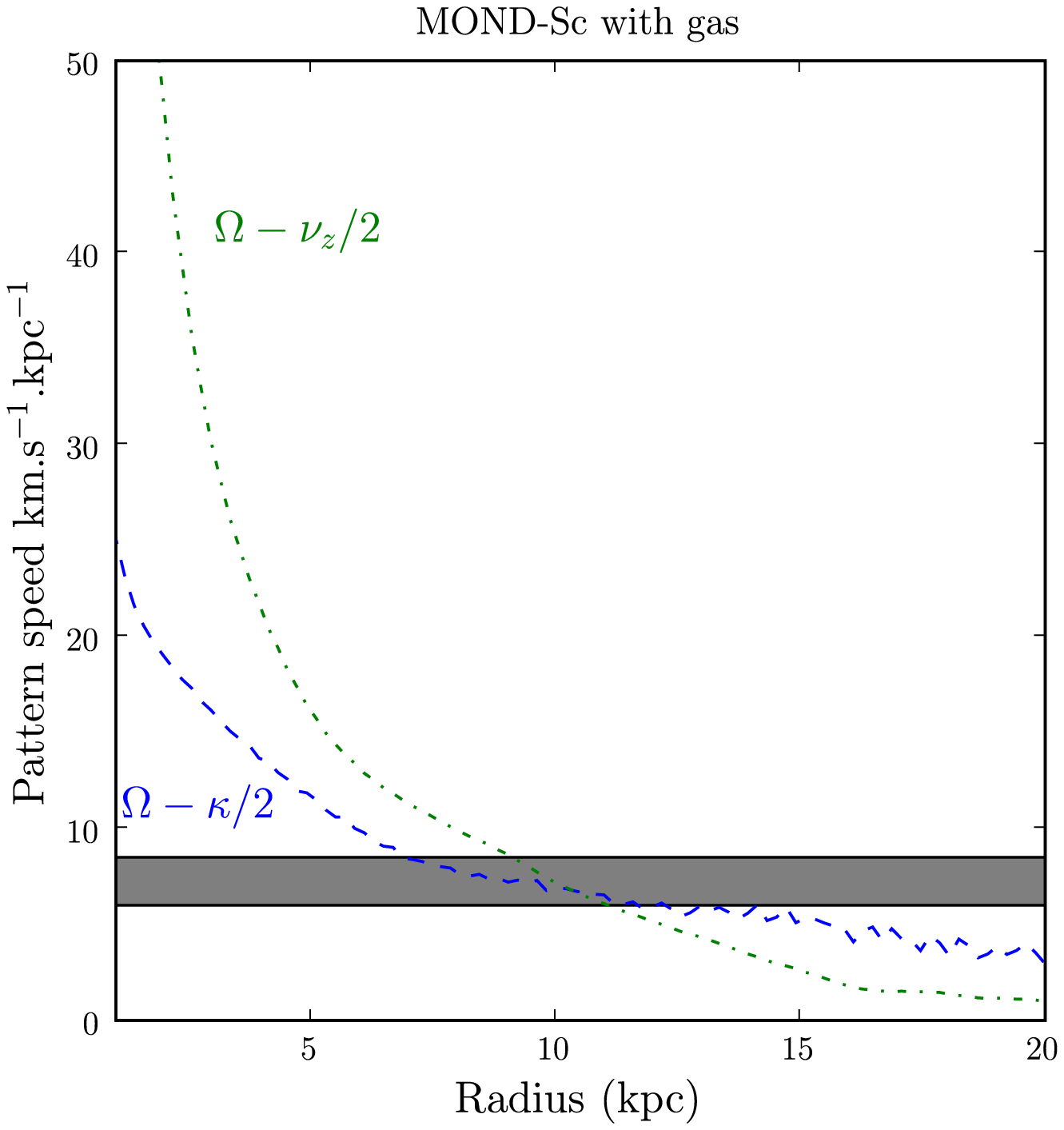}}

\caption{Localisation of the peanut resonance for the MOND-Sc galaxy without gas (left) and with gas (right). With gas, the bar pattern speed is about 8 km.s$^{-1}$.kpc$^{-1}$ (the gray region marks the variation of the bar pattern speed during time, see text) so the peanut resonance occurs near 9 kpc, while it appears at 12 kpc without gas.}

\label{fig:rezo_z}
\end{figure}

For the simulations without gas, all the DM-simulations and the MOND-Sa simulation clearly show the buckling, while the peanut looks like a boxy bar for the MOND-Sb and MOND-Sc. In the DM-Sc simulation, the peanut is not symmetric because it is still forming, as the bar formation is slower (Fig. \ref{fig:barstrength}). 
The Fig. \ref{fig:peanut} plots are the stellar component only. The gas responds to the peanut resonance and forms a vertical structure too, but it comes back quite rapidly (1 Gyr) to the galactic plane by collisions. Moreover, by dissipation and angular momentum transfer during the bar formation, the gas peaks in the nucleus and creates a large concentration in the stellar component too. Then the bar pattern speed is higher in simulation with gas and the resonances 
(bar in the plane, peanut vertically) occur closer to the galactic centre (Fig. \ref{fig:rezo_z}). This is why the models with gas appear less extended (like the MOND-Sc).

\begin{figure}
 \centering
 \resizebox{\hsize}{!}{\includegraphics{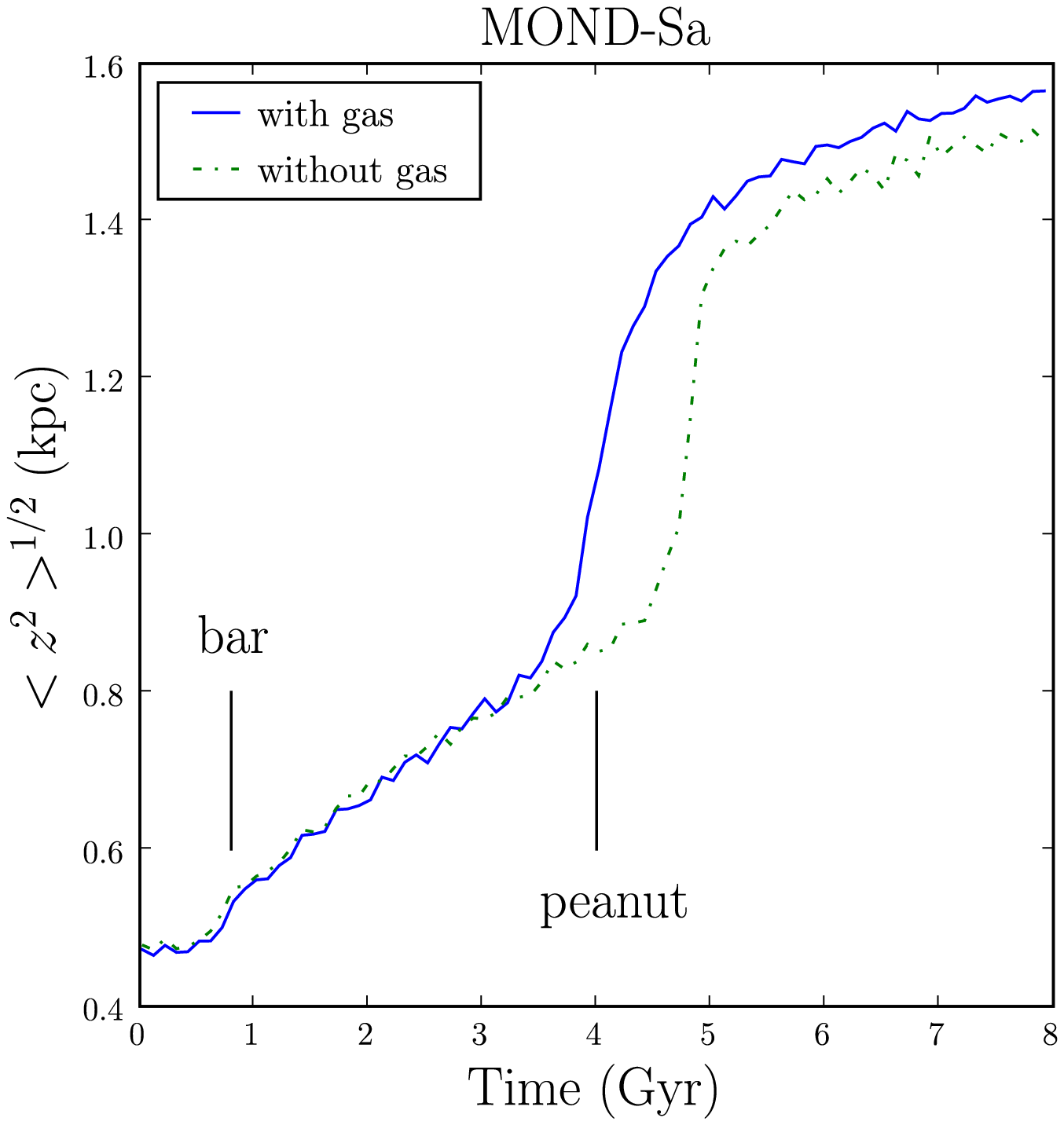}\includegraphics{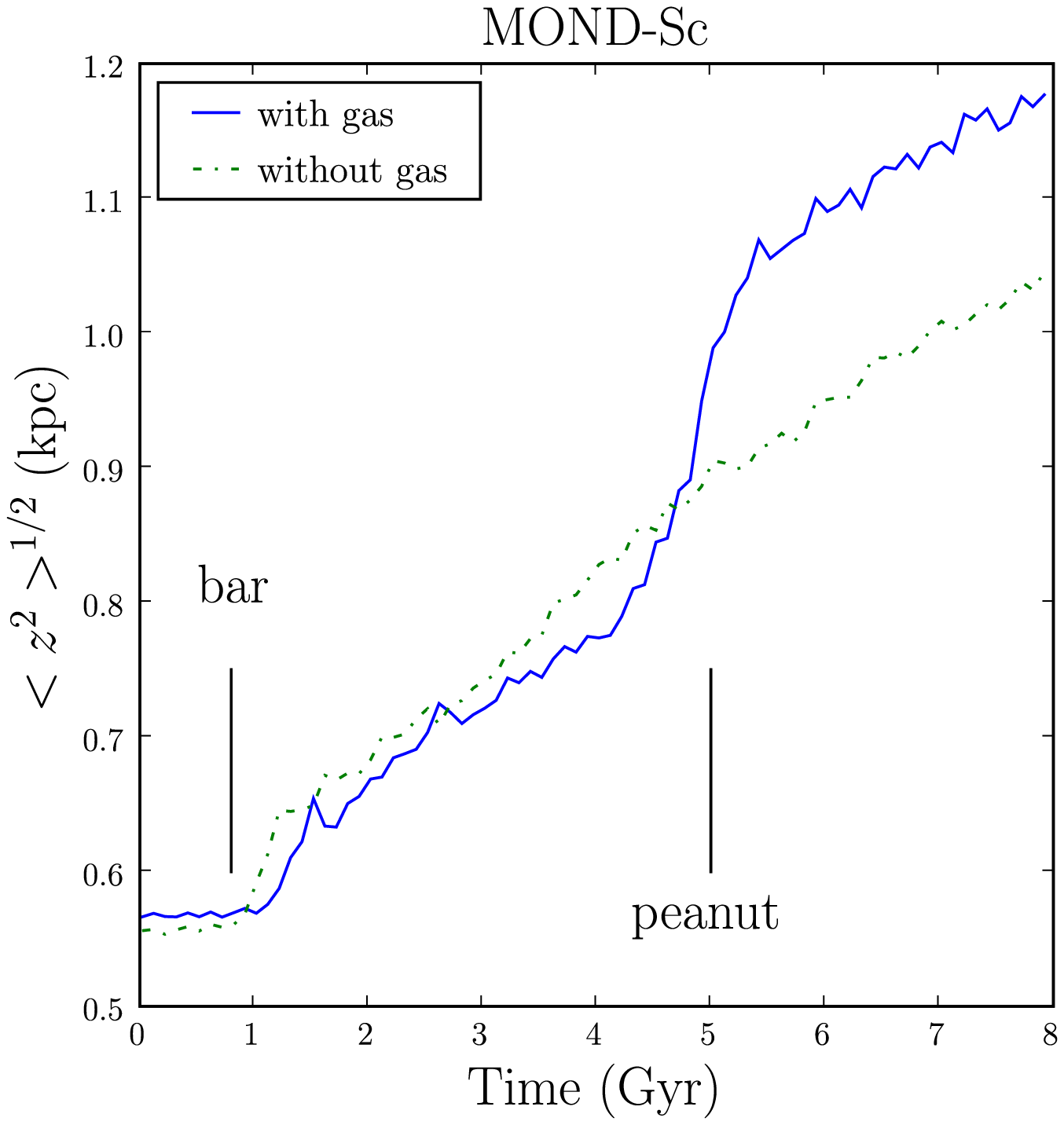}}
 \resizebox{\hsize}{!}{\includegraphics{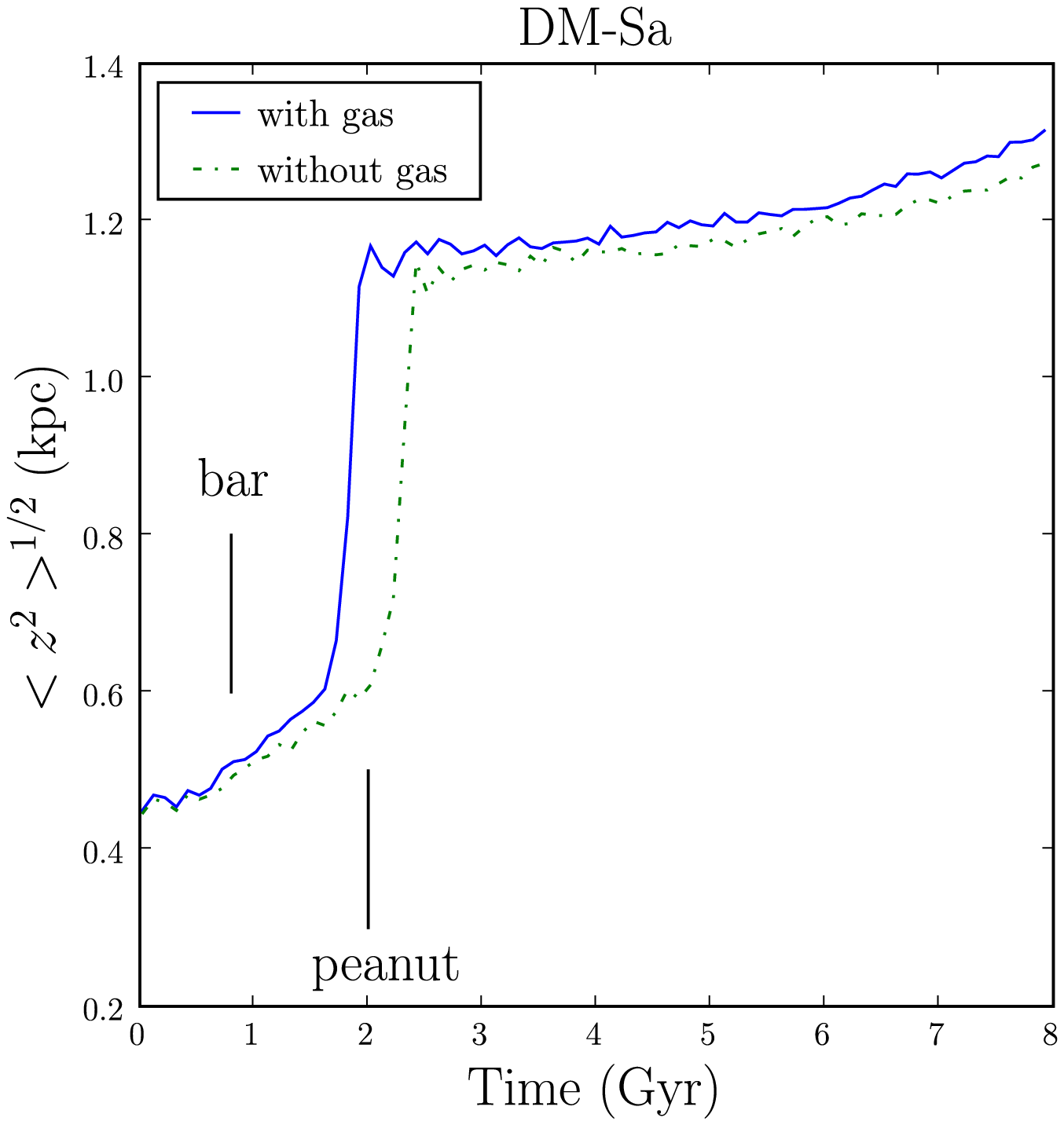}\includegraphics{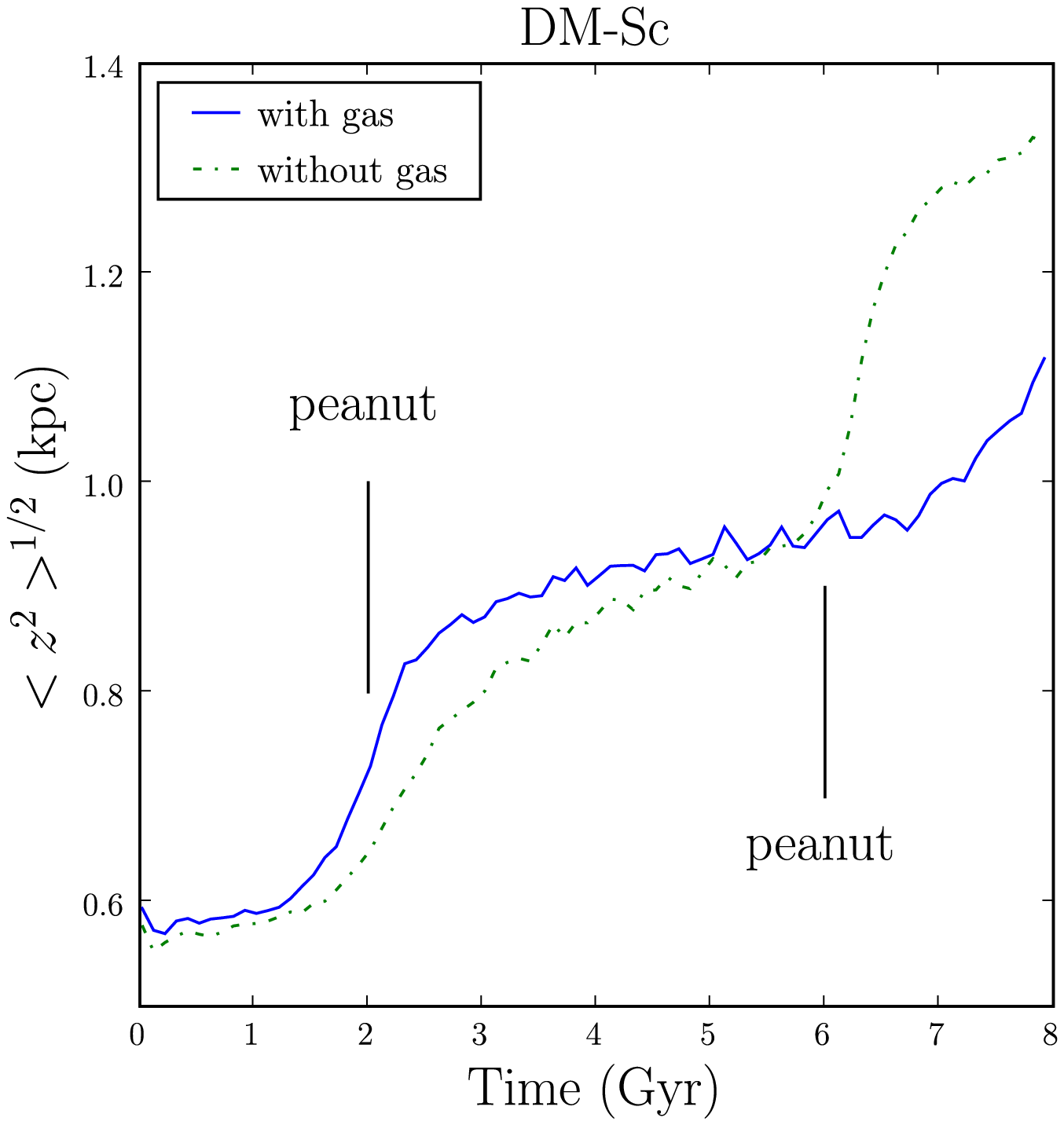}}

\caption{Evolution of the stellar disc thickness, $<z^2>^{1/2}$, for the Sa and Sc galaxies, in the DM and MOND models. The formation of spiral arms and the bar instability heat the disc which thickens progressively. The steps correspond to the peanut resonance when the stellar particles leave suddenly the galactic plane.  }

\label{fig:z2}
\end{figure}

In Fig. \ref{fig:z2}, the thickness is traced by the root mean square of the vertical position of particles inside 15 kpc. For the MOND-Sa, the buckling occurs around 4-5 Gyr, the mean thickness increases from 0.8 kpc to 1.4 kpc, as in the DM simulation, where it appears at 2 Gyr and thickens the disc from  0.6 kpc to 1.2 kpc. In the MOND simulation, the disc continues to heat, while yielding a thick disc (Fig. \ref{fig:peanut}). In the DM simulation, the buckling is shifted outward in the disc while the bar slows down (Fig.\ref{fig:dfric}), more and more particles get out from the galactic plane (after 6 Gyr for the DM-Sa galaxy).

For the MOND-Sc galaxy, the peanut seems to appear only in the gas model at 5 Gyr. The disc thickens from 0.8 kpc to 1. kpc. Without gas, no sudden thickening is measured with this indicator, but the galaxy displays a boxy shape along the long and short axis of the bar. 

In the DM-Sc simulation, when the bar growth rate is slow, two episodes of buckling can be distinguish (Martinez-Valpuesta et al. 2006); one around 3 Gyr and the other at 7 Gyr, the disc thickens from 0.6 kpc, 0.8 kpc to 1.3 kpc.

In these simulations, it can be observed that the peanut resonance occurs sooner in the DM-model than in the MOND-model. On the average, it appears 2-3 Gyr after the beginning of the simulation in the DM model against 4-5 Gyr in the MOND model.

Bars are weakened during the buckling (Berentzen et al. 2007), but this is not the only way. The MOND-Sc simulation with gas does not have a strong peanut, but the bar is weakened because the instability is too violent during a short period: the bar strength increases to 0.4 in less than 1 Gyr. The orbits of particles have no
time to follow regular patterns, and chaos reduces the bar strength.
Equivalent simulations have been performed in 2D and have confirmed that it is not an effect of the vertical motion (TC07) 

Strong peanuts appear preferentially in early-type galaxies (in MOND as well in DM) while late-types galaxies display a boxy bar. This feature should be compared with observation (L\"{u}tticke et al. 2000), but within the statistical uncertainties, it is not possible yet to say whether
some morphological type has more tendency to form a peanut or not.

\subsection{Statistics of bar frequency}

Several groups have estimated bar frequency from observations
(Block et al 2002; Laurikainen et al 2004; Buta et al 2005) and, in the bar strength histogram, found a dearth of weak bars, with an extended wing of strong bars.
In TC07, we estimated the bar frequency in a sample of pure stellar galaxy discs. 
The obtained bar frequency with MOND had more similarities to the observed
histogram, since strong bars are more frequent than 
in the DM model. We have obtained similar diagrams with numerical simulations of galaxies, including the gas behaviour and the star formation. We have also run new simulations of galaxies 
without gas for the sake of comparison with the new choice of the standard $\mu$-function in MOND, and new dark matter halo 
distribution in DM. 

Without gas, the new DM distributions have not changed the global result of the statistical study in the
 bar frequency for pure stellar galaxies. There are still fewer galaxies in MOND with a low bar strength and more galaxies that 
are strongly barred compared to DM. 
This can be understood in terms of the bar formation, which is always faster in MOND than in Newton+DM in these types of galaxies. 

Simulations show that the maximum of bar strength, obtained just after
the first instability, corresponds to the
largest growth rate of the bar, for the same galaxy and varying the Toomre parameter Q. The more unstable the galaxy is (for lower Q values) 
the faster the bar grows, and the higher is the maximum reached for the bar strength. 
When the galaxy is violently unstable, the bar is then weakened by feedback 
and chaotic effects (like the MOND-Sc one).

By adding gas, the bar frequency in the DM model increases and tends to approach 
the MOND bar frequency. The bars occur so quickly in MOND that the timescale of bar formation cannot be shortened, contrary to the Newtonian case. In DM + gas the proportion of low bar strength 
galaxies decreases because of the faster bar formation.
In MOND with gas, the bar formation does not change much due to saturation. In the bar strength evolution, we see 
(Fig. \ref{fig:barstrength}) a slight deviation only for the Sc galaxy (with $7\%$ of gas).
The maximum of the bar strength, in DM and MOND, is the same with and without gas. 
So the part of the histogram with high bar strength does not change anymore.
The difference appears in the middle of the distribution. Instead of having a plateau between $0.1$ and $0.3$ in bar
strength, the distribution is peaked at $0.15$, $0.20$. In addition, 
bars are weakened or destroyed more quickly if there is gas because of gravity torques, and gas inflow.

To summarise, the low bar strength are constrained by the time-scale to form a bar, the strongly barred galaxies correspond to the 
capacity for a galaxy to have a high bar strength after the bar formation. And the middle of the distribution shows the evolution 
of the bars after their formations. Several factors, such as saturation effects in bar formation and bar weakening due to gas inflows, combine to minimise the difference from previous results without gas.

\begin{figure}
  \resizebox{\hsize}{!}{\includegraphics{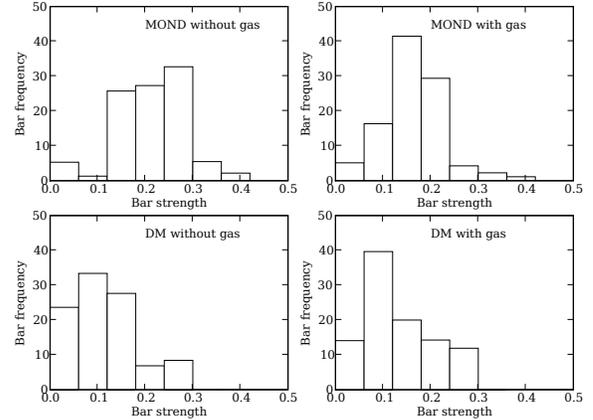}}
     \caption{Histograms of bar strength from the simulated galaxies in the MOND (top) and DM (bottom) models,
without (left) and with gas (right). }
     \label{fig:barfreq}
\end{figure}

\section{Discussion and conclusion}

In this paper, we present new simulations of isolated galaxies in both frameworks, of Newtonian gravity with dark matter and MOND,
to compare their stability and bar frequency, and try to constrain either one by confrontation with observations. 
The present simulations now include the gas dissipation in the disc and star formation. For the MOND model, the 
standard $\mu$-function is used instead of the simple one (as in TC07). We will then also discuss the particular effects of varying
$\mu$-functions, in addition to the effect of the gas on galaxy evolution.

First, one remarkable result is that the galaxy evolution in MOND is not affected by the choice of the $\mu$-function. Whatever the 
adopted interpolating function (standard or simple), a galaxy forms a bar in the same timescale, and with the same bar strength. This 
hypothesis that the exact behaviour of $\mu$-function is not important, but the asymptotic behaviour is, was made by Milgrom and 
had not yet been tested with numerical simulations.
It could be interesting to extend this test with the entire existing family of $\mu$-functions. We have not made this comparison to reduce 
the number of parameters and make the comparison with the Newtonian gravity more clear.

In our simulations, we constrain the initial  galaxies to have the same rotation curve in both MOND and Newtonian dynamics. 
We also choose the same velocity dispersion in the discs for the two dynamics.
If the choice of the standard or the simple $\mu$-function does not affect the MOND model,
it, however, significantly changes (by a factor 2) the mass of the ``phantom'' dark matter halo inside 
the visible radius required in the Newtonian dynamics, which stabilises the galaxy disc against bar formation.
 Since there is now less dark matter in the Newton models, the bar pattern speeds are no longer slowed
down in large proportions, as was the case on our previous work (TC07). This is now more compatible with
observations.

As expected, the gas makes the disc more unstable, decreasing the timescale to form a bar especially in Newton+DM, where it was
previously longer than in MOND. With gas, the histogram 
of bar frequency for the DM model tends to be more similar to the MOND one. 
Still galaxies in MOND are more often barred, and are also more strongly barred. 
 The bar frequency histogram is compatible with the data in both models, although the 
MOND results give a better fit.

 The introduction of gas accelerates the bar formation and, at the same time the 
peanut formation. Contrary to what was found by Berentzen et al (2007), the introduction of gas does
not damp  the buckling instability, in the models considered here (both in MOND and DM).  In MOND, without gas, the mass concentration in the
Sc-model is lower in the centre, and the position of the vertical instability occurs at a larger radius,
leading to a less conspicuous box-peanut shape in the edge-on view.

The bar gravity torques drive the gas inwards inside corotation (e.g. Buta \& Combes 1996), and outwards outside,
producing gas accumulation in resonant rings. Young stars are formed in these rings, which are then
conspicuous in optical images. 
We show that MOND can reproduce as well as Newtonian gravity these resonant structures,
 rings and pseudo-rings, and inner and outer Linblad resonances, or ultra-harmonic resonance,
near corotation.

This work constitutes a more realistic step from the pure stellar disc simulations, for isolated galaxies.
Future simulations (Tiret \& Combes 2007b, Tiret \& Combes in prep.) will deal with galaxy interaction, and 
dynamical friction, where gas dissipation 
and star formation plays a significant role, and which are a challenge for MOND.

\begin{acknowledgements}
We thank the referee for very helpful remarks.
We are grateful to Y. Revaz for enlightening discussions. 
Simulations in this work have been carried out with the IBM-SP4 of the CNRS computing centre, at IDRIS (Orsay, France)
\end{acknowledgements}


\begin{thebibliography}{}
\bibitem{} Athanassoula, E. : 2002, ApJ  569, L83
\bibitem{} Avila-Reese, V., Carrillo, A., Valenzuela, O.,  Klypin, A.: 2005, MNRAS, 361, 997
\bibitem{} Berentzen, I., Shlosman, I., Martinez-Valpuesta, I., Heller, C. H.: 2007 ApJ, 666, 189
\bibitem{} Brada, R., Milgrom, M.: 1999, ApJ 519, 590
\bibitem{} Block, D. L., Bournaud, F., Combes, F., Puerari, I., Buta, R.: 2002 A\&A 394, L35
\bibitem{} Bournaud, F., Combes, F.: 2002, A\&A 392, 83
\bibitem{} Bournaud, F., Combes, F., Semelin, B.: 2005, MNRAS 364, L18
\bibitem{} Buta, R. \& Combes, F.: 1996, Fundamentals of Cosmic Physics, Volume 17, pp. 95-281
\bibitem{} Buta, R., Vasylyev, S., Salo, H., Laurikainen, E.: 2005 AJ 130, 506
\bibitem{} Ciotti, L. , Binney, J.: 2004, MNRAS 351, 285
\bibitem{} Col\'{\i}n, P., Valenzuela, O., Klypin, A.: 2006, ApJ, 644, 687
\bibitem{} Combes, F. , Gerin, M.: 1985, A\&A 150, 327
\bibitem{} Debattista, V. P., Sellwood, J. A.: 1998, ApJ, 493, 5D
\bibitem{} Debattista, V. P., Sellwood, J. A.: 2000, ApJ, 543, 704
\bibitem{} Dubinski, J., Mihos, J.C., Hernquist, L.:  1996, ApJ  462, 576
\bibitem{} Hernquist, L., Weinberg, M. D.: 1992, ApJ 400, 80
\bibitem{} Laurikainen, E., Salo, H., Buta, R., Vasylyev, S.: 2004 MNRAS 355, 1251
\bibitem{} Little, B., Carlberg, R. G.: 1991, MNRAS 251, 227
\bibitem{} L\"{u}tticke R., Dettmar R.-J., Pohlen M.: 2000, A\&AS, 145, 405
\bibitem{} Martinez-Valpuesta, I.; Shlosman, I.; Heller, C.: 2006 ApJ 637, s214M 
\bibitem{} Mayer, L., Wadsley, J.: 2004, MNRAS, 347, 277
\bibitem{} Milgrom, M.: 1983, ApJ 270, 371
\bibitem{} Milgrom, M.: 1989, ApJ 338, 121
\bibitem{} Nipoti, C., Ciotti, L., Binney, J., Londrillo, P.: arXiv, 0802.1122
\bibitem{} O'Neill, J. K., Dubinski, J.: 2003, MNRAS, 346, 251
\bibitem{} Ostriker J., Peebles, P.J.E.: 1973, ApJ 186, 46
bibitem{} Piner B.G., Stone J.M., Teuben P.J.: 1995, ApJ 449, 508
\bibitem{} Press W.H., Teukolsky S.A., Vetterling W.T., Flannery B.P., Numerical Recipes in Fortran 77, second edition, Cambridge University Press, 1992
\bibitem{} Regan, M. W., Teuben, P.: 2003 ApJ 582, 723
\bibitem{} Sanders R.H., McGaugh S.S.: 2002, ARAA 40, 263
\bibitem{} Schwarz M.P.: 1984, MNRAS 209, 93
\bibitem{} Sellwood, J. A.; Debattista, V. P.: 2006 ApJ, 639, 868
\bibitem{} Tiret, O., Combes, F.: 2007a, A\&A 464, 517, TC07
\bibitem{} Tiret, O., Combes, F.: 2007b, arXiv, 0712.1459
\bibitem{} Valenzuela, O., Klypin, A.: 2003, MNRAS, 345, 406
\bibitem{} Weinberg, M. D.: 1985, MNRAS, 213, 451
\bibitem{} Zhao, H.S., Famaey, B.: 2006, ApJ, 638, L9
\end{thebibliography}
\end{document}